\newcommand{\exN}{10}
\newcommand{\exAL}{40}

\newcommand{\Hi}{\ion{H}{1}}
\newcommand{\HEi}{\ion{He}{1}}
\newcommand{\HEii}{\ion{He}{2}}
\newcommand{\Ciii}{\ion{C}{3}}
\newcommand{\Civ}{\ion{C}{4}}
\newcommand{\Nii}{\ion{N}{2}}
\newcommand{\Niii}{\ion{N}{3}}
\newcommand{\Niv}{\ion{N}{4}}
\newcommand{\Oi}{\ion{O}{1}}
\newcommand{\NEiii}{\ion{Ne}{3}}
\newcommand{\NEv}{\ion{Ne}{5}}
\newcommand{\ALii}{\ion{Al}{2}}
\newcommand{\ALiii}{\ion{Al}{3}}
\newcommand{\SIii}{\ion{Si}{2}}
\newcommand{\FEii}{\ion{Fe}{2}}

\documentclass[twocolumn,doublespacing]{aastex631}
\usepackage{enumitem}

\shorttitle{A Low-mass ONeMg WD behind V1405 Cassiopeiae}
\shortauthors{Taguchi et al.}
\graphicspath{{./}{figures/}}

\begin{document}

\title{Spectra of V1405 Cas at the very beginning indicate a low-mass ONeMg white dwarf progenitor}

\correspondingauthor{Kenta Taguchi}
\email{kentagch@kusastro.kyoto-u.ac.jp, taguchi.kenta.34z@kyoto-u.jp}

\author[0000-0002-8482-8993]{Kenta Taguchi}
\affiliation{Department of Astronomy, Kyoto University, Kitashirakawa-Oiwake-cho, Sakyo-ku, Kyoto 606-8502, Japan}

\author[0000-0003-2611-7269]{Keiichi Maeda}
\affiliation{Department of Astronomy, Kyoto University, Kitashirakawa-Oiwake-cho, Sakyo-ku, Kyoto 606-8502, Japan}

\author[0000-0003-0332-0811]{Hiroyuki Maehara}
\affiliation{Subaru Telescope Okayama Branch Office, National Astronomical Observatory of Japan, 3037-5, Honjou, Kamogata, Asakuchi, Okayama 719-0232, Japan}

\author[0000-0001-8813-9338]{Akito Tajitsu}
\affiliation{Subaru Telescope Okayama Branch Office, National Astronomical Observatory of Japan, 3037-5, Honjou, Kamogata, Asakuchi, Okayama 719-0232, Japan}

\author[0000-0001-9456-3709]{Masayuki Yamanaka}
\affiliation{Okayama Observatory, Kyoto University, 3037-5 Honjo, Kamogata-cho, Asakuchi, Okayama 719-0232, Japan}
\affiliation{Amanogawa Galaxy Astronomy Research Center (AGARC), Graduate School of Science and Engineering, Kagoshima University, Kagoshima 890-0065, Japan}

\author[0000-0002-5756-067X]{Akira Arai}
\affiliation{Subaru Telescope, National Astronomical Observatory of Japan, 650 North A'ohoku Place, Hilo, HI 96720, USA}

\author[0000-0002-6480-3799]{Keisuke Isogai}
\affiliation{Okayama Observatory, Kyoto University, 3037-5 Honjo, Kamogata-cho, Asakuchi, Okayama 719-0232, Japan}
\affiliation{Department of Multi-Disciplinary Sciences, Graduate School of Arts and Sciences, The University of Tokyo, 3-8-1 Komaba, Meguro, Tokyo 153-8902, Japan}

\author{Masaaki Shibata}
\affiliation{Department of Astronomy, Kyoto University, Kitashirakawa-Oiwake-cho, Sakyo-ku, Kyoto 606-8502, Japan}

\author[0000-0003-0196-3936]{Yusuke Tampo}
\affiliation{Department of Astronomy, Kyoto University, Kitashirakawa-Oiwake-cho, Sakyo-ku, Kyoto 606-8502, Japan}

\author{Naoto Kojiguchi}
\affiliation{Department of Astronomy, Kyoto University, Kitashirakawa-Oiwake-cho, Sakyo-ku, Kyoto 606-8502, Japan}

\author{Daisaku Nogami}
\affiliation{Department of Astronomy, Kyoto University, Kitashirakawa-Oiwake-cho, Sakyo-ku, Kyoto 606-8502, Japan}

\author{Taichi Kato}
\affiliation{Department of Astronomy, Kyoto University, Kitashirakawa-Oiwake-cho, Sakyo-ku, Kyoto 606-8502, Japan}



\begin{abstract}
  The lowest possible mass of ONeMg white dwarfs (WDs) has not been clarified despite its importance in the formation and evolution of WDs.
  We tackle this issue by studying the properties of V1405 Cas (Nova Cassiopeiae 2021), which is an outlier given a combination of its very slow light-curve evolution and the recently reported neon-nova identification.
  We report its rapid spectral evolution in the initial phase, covering 9.88, 23.77, 33.94, 53.53, 71.79, and 81.90 hours after the discovery.
  The first spectrum is characterized by lines from highly-ionized species, most noticeably {\HEii} and {\Niii}.
  These lines are quickly replaced by lower-ionization lines, e.g., {\Nii}, {\SIii}, and {\Oi}.
  In addition, {\ALii} (6237 {\AA}) starts emerging as an emission line at the second epoch.
  We perform emission-line strength diagnostics, showing that the density and temperature quickly decrease toward later epochs.
  This behavior, together with the decreasing velocity seen in H$\alpha$, H$\beta$, and {\HEi}, indicates that the initial nova dynamics is reasonably well described by an expanding fireball on top of an expanding photosphere.
  Interestingly, the strengths of the {\Niii} and {\ALii} indicate large abundance enhancement, pointing to an ONeMg WD progenitor as is consistent with its neon-nova classification.
  Given its low-mass nature inferred by the slow light-curve evolution and relatively narrow emission lines, it provides a challenge to the stellar evolution theory that predicts the lower limit of the ONeMg WD mass being $\sim$ 1.1 $M_\odot$.
\end{abstract}



\section{Introduction}
\label{sec:intro}

The classical nova is an explosion on the surface of an accreting white dwarf (WD) caused by the thermonuclear runaway, when the mass of the accumulated hydrogen-rich envelope has reached a critical mass \citep{Starrfield_1972}.
The initial rapid brightening (by 5--13 magnitudes) in optical wavelengths typically takes place in less than three days.
It is then followed by the pre-maximum halt and the final rise of about two magnitudes to the optical maximum, which takes another $\sim$ 1--2 days for fast novae or several weeks for the slowest novae (see \citealt{Bode_Evans_2008} for the observational properties of novae).

Theoretically, the temperature is considered to be high in the initial phase \citep{Hillman_2014,Kato_eROSITA}, which provides a different condition for spectral formation as compared to the later, well-observed epochs.
Therefore, spectroscopic observations in such an infant phase are expected to provide new insight into the natures of the WD progenitor and the ejecta dynamics just after the initiation of the thermonuclear runaway.
However, it is an extremely challenging observation, given the short time scale associated with the initial brightening; except for some symbiotic novae or extremely slow novae, e.g., Gaia22alz \citep{Aydi_2023,ATel_15270}, only a single example, the recurrent nova T Pyx in its 2011 outburst, has been reported for such a prompt spectroscopic observation \citep{Arai_2015}.

One potential benefit that could be added by such a prompt spectroscopic observation, as we demonstrate in the present work, is the nature of a progenitor WD.
Depending on their internal chemical composition, WDs are largely divided into two types, CO WDs and ONeMg WDs.
However, it is usually difficult to determine their nature either as a CO WD or ONeMg WD by observing `static' WDs, since the internal composition is hidden by a layer of H or He on the surface.
A nova provides an interesting alternative opportunity; the nature of the underlying WD can be inferred either as a CO WD or ONeMg WD, through the spectral identification, i.e., `CO nova' for the former and `neon nova' or `ONe nova' (or `ONeMg nova') for the latter, where the classification as a neon nova is based on the overabundance of neon derived from neon lines, especially [{\NEiii}] and [{\NEv}] in the decline (coronal) phase \citep[e.g.,][]{Bode_Evans_2008}.
Further, it is possible to estimate the WD mass behind a nova.
The `speed class', i.e., fast or slow, is conventionally defined by the decay speed in the decline phase after the maximum light\footnote{
  In this article, we follow \cite{Payne-Gaposchkin_1964} to determine the speed class.
  For Galactic novae before 2010, we use decline times given by \cite{Strope_2010}.
}.
It is generally accepted that novae with faster (slower) declines in the light curves are associated with heavier (lighter) WDs\footnote{
  While the WD mass is the main controlling factor, there are also other factors that affect the nova speed class \citep[e.g., the accretion rate from the secondary star;][]{Yaron_2005}.
} \citep{Kato_1994,Hachisu_2010}.

Especially interesting in this context is a mass range for ONeMg WDs.
Compared to CO WDs, ONeMg WDs are predicted to be generally more massive, as they are end products of more massive stars.
Theoretically, the minimal mass of ONeMg WDs has long been studied by stellar evolution simulations \citep[e.g.,][]{Nomoto_1984,Iben_1985,Nomoto_1987}.
According to a recent study \citep{Lauffer_2018}, it is $\sim$ 1.1 $M_{\odot}$.
However, there are some observational works that challenge this picture, and the minimal mass of ONeMg WDs has not been clarified.
For example, \cite{Kepler_2016} reported a discovery of an ONeMg WD with 0.56 $\pm$ 0.09 $M_{\odot}$.
Observations of novae can form potentially important data sets to tackle this issue, as the WD composition and the mass can be connected for individual objects.

Neon novae almost exclusively belong to fast or very fast novae\footnote{
  The opposite is not the case, i.e., there are many CO novae showing fast light curves, e.g., the recurrent novae V3890 Sgr \citep{Orio_2020} and RS Oph \citep{Mikolajewska_2017}.
} \citep{Bode_Evans_2008} and are therefore associated with a massive ONeMg WD.
This picture was however challenged by the discovery of a slow {\em and} neon nova -- \cite{Takeda_2018} concluded the very slow nova V723 Cas (Nova Cassiopeiae 1995) is a neon nova, hosting a WD of only $\sim$ 0.6 $M_{\odot}$.
To further investigate the mass range of ONeMg WDs (especially toward the low-mass end), an additional support of an ONeMg WD progenitor for a slow nova that shows potential signatures of a neon nova will provide an important step.

In the present work, we show the result of our follow-up spectroscopy of the very slow nova V1405 Cas which was recently reported to be a neon nova based on its spectra in the decay phase \citep{ATel_15796}, i.e., the second example of a very slow nova categorized as a neon nova.
We present very early-phase spectral evolution soon after the discovery, including a spectrum taken during the initial rise, and report the detection of {\ALii} $\lambda$6237 and {\ALiii} $\lambda\lambda$ 5697/5723.
We perform line-strength diagnostics that shows overabundances of aluminum, suggesting that the progenitor WD is indeed an ONeMg WD despite its likely low mass nature in view of the slow light-curve evolution.
This paper is structured as follows.
We describe the basic properties of V1405 Cas in \S\ref{sec:v1405cas}.
Our observations and data reductions are shown in \S\ref{sec:spectra}, where we also describe overall spectral evolution.
In \S\ref{sec:1zone}, we diagnose the physical condition of the system from the line strengths.
In \S\ref{subsec:discussion:v1405cas}, we discuss the possibility that V1405 Cas is a low-mass ONeMg WD.
In \S\ref{subsec:discussion:past_neon_novae}, we discuss the possible correlation between aluminum overabundance and neon novae.
In \S\ref{subsec:discussion:6237}, we discuss how {\ALii} $\lambda$6237 can be used as the earliest indicator to identify neon nova candidates that may help coordinate intensive observations for similar events in the future.
We conclude this paper in \S\ref{sec:conclusion}.

\section{Basic Properties of V1405 Cas}
\label{sec:v1405cas}

V1405 Cas (= Nova Cassiopeiae 2021 = PNV J23244760+6111140) was discovered as a transient candidate by Yuji Nakamura at 9.6 mag (unfiltered) on 2021-03-18.4236 UT (JD 2459291.9236)\footnote{
  \url{http://cbat.eps.harvard.edu/unconf/followups/J23244760+6111140.html}\label{foot:CBAT}
}.
Hereafter, we denote $t$ as the time measured from the discovery time by Nakamura\footnote{
  While S. Korotkiy et al. reported to CBAT{\footnotesize $^{\text{\ref{foot:CBAT}}}$} that the system was seen at 13.53 $\pm$ 0.10 mag (unfiltered) on 2021-03-14.71017 ($t$ = $-$3.7134 days), by the  New Milky Way (NMW) survey (\url{http://scan.sai.msu.ru/nmw/}), this is far below the maximum magnitude.
}.
It was then confirmed as a classical nova \citep{ATel_14471,ATel_14472}, who reported the first spectrum shown in the present paper.

The position of the nova coincides with a Gaia Data Release 3 \citep{Gaia_2016b,Gaia_2022k} source (2015451512907540480), which is characterized by an average G magnitude = 15.36 and a parallax p = 0.5776103855139701 ${\pm}$ 0.025421817 mas.
Hereafter, we adopt d = 1.73 $_{-\text{0.07}}^{+\text{0.08}}$ kpc as the distance to the nova, as inferred from the Gaia parallax.
Taking the color\footnote{
  In addition to the Gaia Data Release 3, the Pan-STARRS1 Database \citep{2020ApJS..251....7F} also records a source PS1  181423511988865624 at the position consistent with the nova.
  The PS1 magnitudes were g = 15.67, r = 15.46, i = 15.44, z = 15.36, and y = 15.32.
  \label{foot:PS1}
} and Gaia distance into account, the progenitor is most likely a nova-like variable \citep{ATel_14472,2021arXiv210613907S}.
Recently, \citet{2021arXiv210613907S} derived its orbital period of 0.1883907 $\pm$ 0.0000048 days from the TESS data \citep{2015JATIS...1a4003R} taken before the outburst \citep[see also the Czech Variable Star Catalogue\footnote{\url{http://var2.astro.cz/czev.php?id=3217}.};][]{CzeV}.

V1405 Cas initially showed He/N type spectra, but after a month it started to show {\FEii} type spectra \citep{ATel_14577,ATel_14614}.
After that, following the luminosity decrease, it showed He/N type spectra again \citep{ATel_14622}.
Finally, based on the detection of overwhelming neon lines at $t$ = +618 days, it was classified as a neon nova \citep{ATel_15796}.

\begin{figure}[htb]
  \centering
  \includegraphics[width=0.9\hsize]{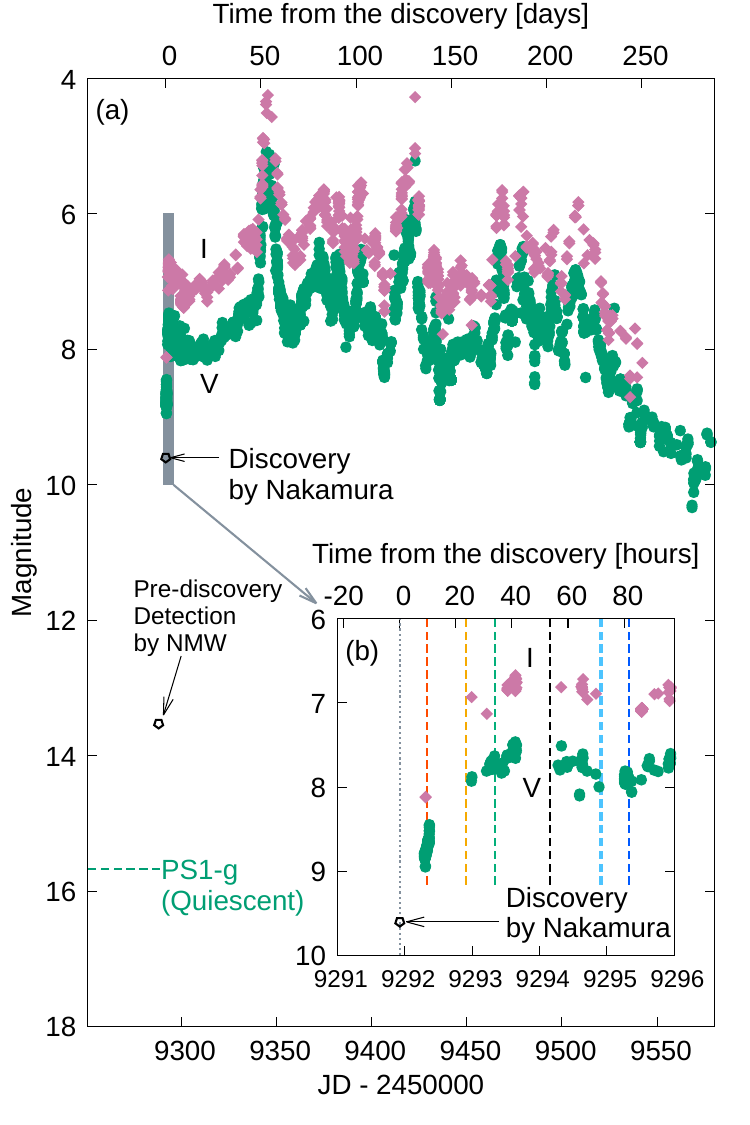}
  \caption{
    The light curves of V1405 Cas.
    (a) V and I magnitudes provided by AAVSO are shown by the green-filled circles and magenta-filled diamonds, respectively.
    The black open pentagons show the pre-discovery magnitude by the NMW survey and the discovery magnitude by Nakamura.
    (b) An enlarged view of the initial phase (2459291 $<$ JD $<$ 2459296, the shaded region in the panel a.
    The dashed lines show the epochs of spectroscopic observations listed in Table \ref{tab:data_set}, with the same  color coordinate used to show our spectra in Figures \ref{fig:v1405cas-spec} and \ref{fig:v1405cas-spec-vel}.
  }
  \label{fig:lightcurve}
\end{figure}

From the light-curve evolution, V1405 Cas can be classified as a very slow nova.
Since May 2021, V1405 Cas exhibited a seven-month plateau during which it repeatedly showed increases and decreases in its magnitude, whose amplitude reached 2--3 magnitudes (Figure \ref{fig:lightcurve}).
Such a light curve is similar to those of slow novae, including V723 Cas \citep{Iijima_1998,Iijima_2006}, HR Del \citep{Rafanelli_1978}, V1819 Cyg \citep{Whitney_1989}, and V5558 Sgr \citep{Tanaka_2011,Poggiani_2008}.

\section{Spectral Properties of V1405 Cas in the Very Early Phase}
\label{sec:spectra}

\subsection{Observations and Data Reduction}
\label{subsec:spectra:observation}

We performed optical spectroscopy of V1405 Cas using the fiber-fed integral field spectrograph (KOOLS-IFU; \citealt{KOOLS-IFU}) on the 3.8-m Seimei telescope \citep{Seimei} at Okayama Observatory, Kyoto University.
The data were taken under the programs 21A-N-CT06 and 21A-K-0017.
We used VPH-blue, VPH-red, and VPH683 grisms, which cover 4000 -- 8900 {\AA}, 5800 -- 10200 {\AA}, and 5800 -- 8000 {\AA} with the spectral resolution of $\sim$ 500, $\sim$ 800, and $\sim$ 2000, respectively.
%

Bias subtraction was performed using IRAF\footnote{
  IRAF is distributed by the National Optical Astronomy Observatories, which are operated by the Associations of Univesities for Research in Astronomy, Inc., under cooperative agreement with the National Science Foundation.
} in the standard manner.
The gain correction was performed using software developed for KOOLS-IFU data.
Spectra for all the fibers were reduced using the Hydra package \citep{hydra1,hydra2}, within which we used Hg, Ne, and Xe arc lamps for wavelength calibration.
The typical accuracy of our wavelength calibration is $\lesssim$ 0.5 {\AA} for VPH-blue and VPH-red, and $\lesssim$ 0.1 {\AA} for VPH683.
The characteristics in the optical properties for different fibers are corrected for by using twilight flat frames.
For each frame, the sky subtraction was performed using the fibers in which the contribution from the target is negligible.
After the sky subtraction, the fibers for V1405 Cas were selected and combined.
We performed this spectrum combination separately in different wavelength ranges to correct for the atmospheric dispersion.
For all the three grisms, the patterns of the wavelength dependence of count values specific to KOOLS-IFU were corrected for, with the standard star HR7596.


Additionally, a high-resolution spectrum of V1405 Cas was obtained using the High Dispersion Spectrograph \citep[HDS;][]{Noguchi_2002} attached to the 8.2 m Subaru Telescope on 2021 March 20 ($t$ = 53.53 hours).
The configuration was set to cover the wavelength region from 4030 to 6780 {\AA}.
We used 0.45 arcsec slit width, resulting in the spectral resolution of R $\simeq$ 80,000. For the wavelength calibration, we used Th-Ar comparison lamp with a typical error of $<$ 10$^{-\text{3}}$ {\AA}.
Extraction of the spectrum in each order was carried out using the IRAF software in a standard manner \citep[e.g.,][]{Tajitsu_2010}.
The blaze functions for some echelle orders are difficult to derive directly, when broad lines are present within the order.
We first created the blase functions for orders not containing such broad line features.
Then, for the remaining orders, we interpolate the blaze functions from the orders that are located in the neighboring places in the 2D image, as they should share the similar blaze functions.
The resulting blaze functions are then used to connect the spectra at different echelle orders, resulting in the final single flux-normalized spectrum.

The journal of our spectroscopic observations is shown in Table \ref{tab:data_set}.
The epochs of our spectroscopic observations are indicated in Figure \ref{fig:lightcurve}.
We emphasize that the initial brightening had not finished when we performed our first spectroscopy, given V = 8.460 on 2021-03-18.8465 ($t$ = +10.24 hours) obtained by our simultaneous photometric observation using the 0.4-m telescope at Kyoto University \citep{ATel_14472}.

\begin{table*}[t]
  \caption{The log of our spectroscopic observations. $t$ is the time relative to the discovery time by Nakamura (JD = 2459291.9236).}
  \label{tab:data_set}
  \centering
  \begin{tabular}{ccccccc}
    \hline
    Start & End & JD-Middle & $t$ [hours] & Airmass & \multicolumn{2}{c}{Total Exposure Time [s]}\\
    &&&& (start -- end) & (VPH-blue, red, 683) & (HDS)
    \\
    \hline
    2021-03-18.8196 & 2021-03-18.8509 & 2459292.335 & +9.88 & 2.542 -- 2.138 & 1680, 180, 240 & --
    \\
    2021-03-19.4095 & 2021-03-19.4187 & 2459292.914 & +23.77 & 2.566 -- 2.704 & 120, 180, 0 & --
    \\
    2021-03-19.8242 & 2021-03-19.8514 & 2459293.338 & +33.94 & 2.434 -- 2.098 & 1080, 150, 300 & --
    \\
    2021-03-20.6534 & 2021-03-20.6565 & 2459294.155 & +53.53 & 3.6396 -- 3.5286 & -- & 180
    \\
    2021-03-21.4123 & 2021-03-21.4178 & 2459294.915 & +71.79 & 2.701 -- 2.789 & 150, 120, 0 & --
    \\
    2021-03-21.8318 & 2021-03-21.8378 & 2459295.335 & +81.90 & 2.259 -- 2.162 & 180, 150, 180 & --
    \\
    \hline
  \end{tabular}
\end{table*}

\begin{figure*}[htb]
  \centering
  \includegraphics[width=0.92\hsize]{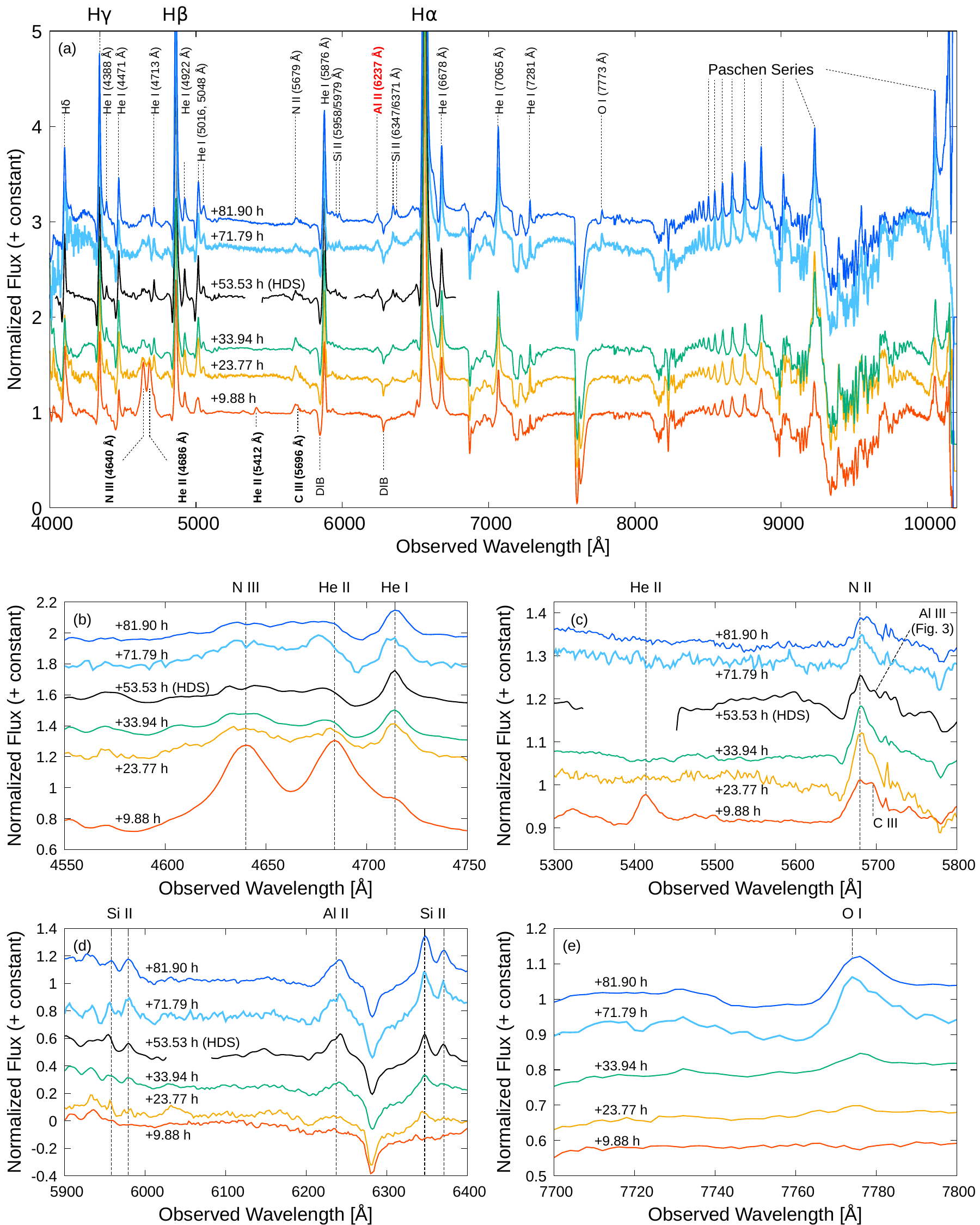}
  \caption{
    Spectra of V1405 Cas by the two low-dispersion grisms (VPH-blue and VPH-red, connected at 6050 {\AA}) of KOOLS-IFU on $t$ = +9.88, 23.77, 33.94, 71.79 and 81.90 hours, and by HDS (with the resolution artificially decreased to the KOOLS-IFU level, \S\ref{subsec:spectra:observation}) on $t$ = +53.53 hours.
    (a) An overall view of the normalized spectra.
    %
    %
    The lines labeled by the bold black font ({\HEii}, {\Ciii}, and {\Niii}) are observed only at $t$ = +9.88 hours.
    The {\ALii} line labeled by the bold red font is one of the most striking lines discussed in \S\ref{subsec:spectra:species_HDS:6200}, \ref{sec:1zone}, and \ref{sec:discussion}.
    (b, c, d, e) Enlarged views of the spectra shown in panel (a).
  }
  \label{fig:v1405cas-spec}
\end{figure*}

\begin{figure}[htb]
  \centering
  \includegraphics[width=0.9\hsize]{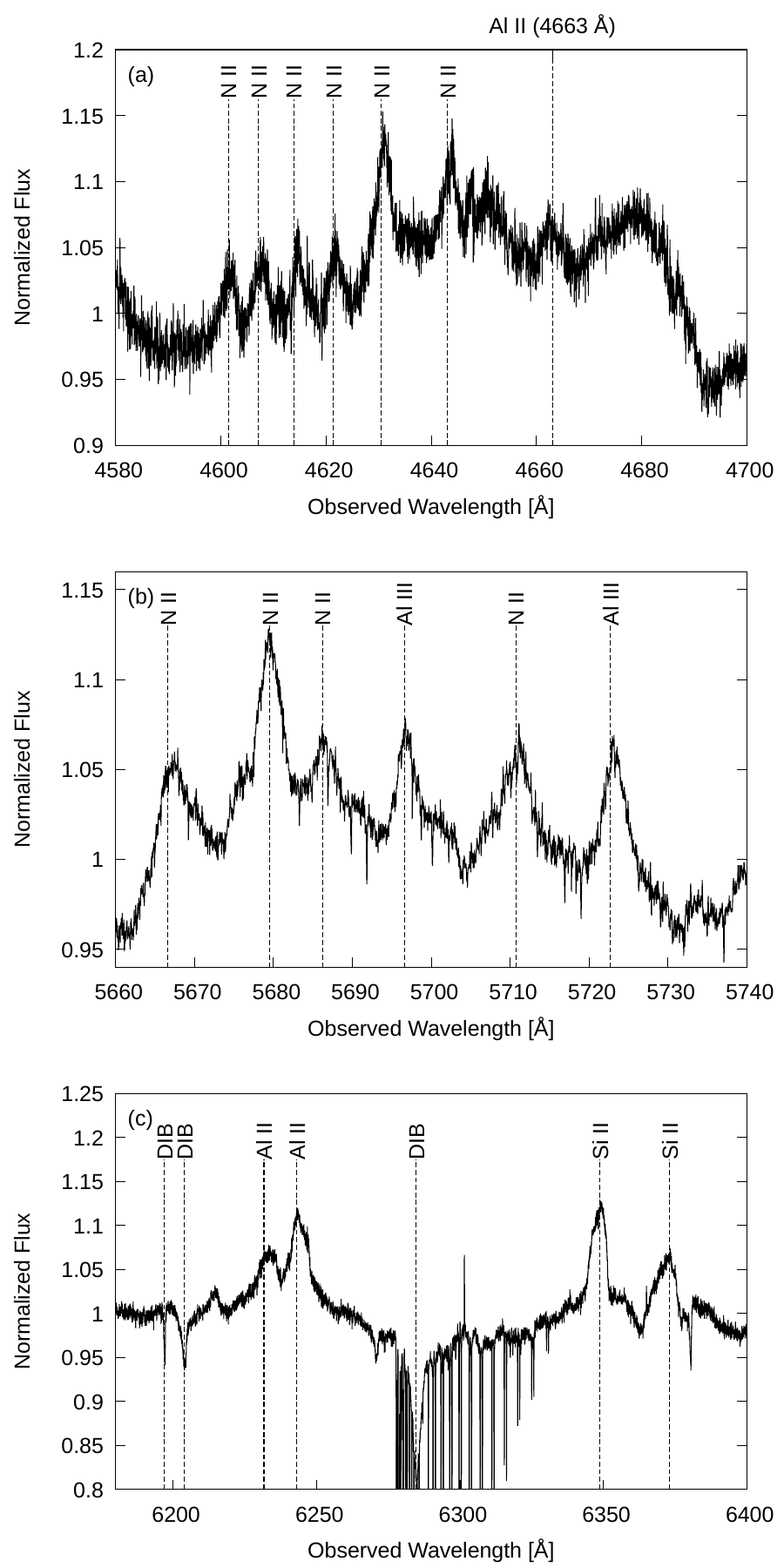}
  \caption{
    Enlarged views of the HDS spectrum on t = +53.53 hours.
    (a) 4580 -- 4700 {\AA} (\S\ref{subsec:spectra:species_HDS:4600}).
    (b) 5660 -- 5740 {\AA} (\S\ref{subsec:spectra:species_HDS:5700}).
    (c) 6180 -- 6400 {\AA} (\S\ref{subsec:spectra:species_HDS:6200}).
  }
  \label{fig:v1405cas-spec-HDS}
\end{figure}

\begin{figure}[htb]
  \centering
  \includegraphics[width=0.9\hsize]{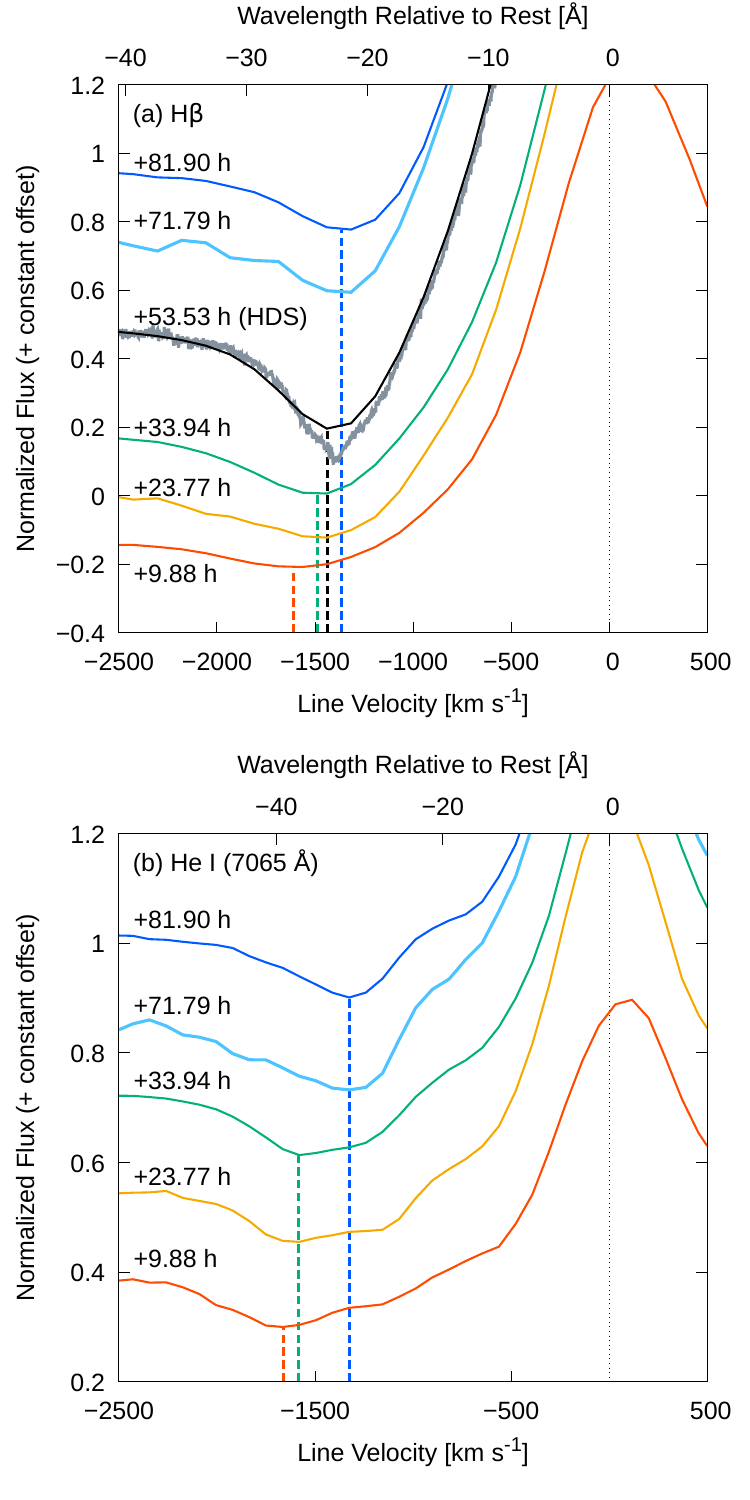}
  \caption{
    Profiles of the H$\beta$ and {\HEi} (7065 {\AA}) lines of V1405 Cas.
    In each panel, the absorption minima at several epochs, measured using the SPLOT task of IRAF, are indicated by dashed lines.
    (a) Spectra around the H$\beta$ line obtained with the VPH-blue grism of KOOLS-IFU and with HDS.
    The absorption minima on $t$ = +9.88, +33.94, and +81.90 hours are at $-$1.6(1), $-$1.4(9), and $-$1.3(6) $\times$ 10$^{\text{3}}$ km s${}^{-\text{1}}$, respectively.
    For the HDS spectrum ($t$ = +53.53 hours), the original high-resolution spectrum (gray) and the resolution-decreased one (black; same as Figure \ref{fig:v1405cas-spec}) are overplotted.
    (b) The same as panel (a), but for the {\HEi} (7065 {\AA}) line obtained using the VPH-red grism.
    The absorption minima on $t$ = +9.88, +33.94, and + 81.90 hours are at $-$1.6(6), $-$1.5(8), and $-$1.3(3) $\times$ 10$^{\text{3}}$ km s$^{-\text{1}}$, respectively.
  }
  \label{fig:v1405cas-spec-vel}
\end{figure}

Figure \ref{fig:v1405cas-spec} shows the overall view of our spectra with the flux scale normalized by the underlying continuum.
Also shown are enlarged views around the wavelength ranges of interest (\S\ref{subsec:spectra:species}).
To assist comparison with the KOOLS-IFU spectra, the HDS spectrum was resolution-decreased by convolving a Gaussian function of $\sigma$ = 100 pixels and then down-sampled.
Figure \ref{fig:v1405cas-spec-HDS} shows enlarged views of the HDS spectrum, supporting line identifications (\S\ref{subsec:spectra:species_HDS}).
Figure \ref{fig:v1405cas-spec-vel} expands the blue-shifted wings of H$\beta$ and {\HEi} (7065\AA).

\subsection{Summary of Evolution of the Spectral Lines}
\label{subsec:spectra:species}
On all the epochs shown in the present paper, the {\Hi} Balmer, Paschen, and {\HEi} lines are detected while {\FEii} lines are not (Figure \ref{fig:v1405cas-spec}a).
The spectra of V1405 Cas, during the initial phase as presented here, thus show characteristics close to those of He/N type novae rather than {\FEii} type ones \citep{Williams_1992,Williams_2012,ATel_14478}\footnote{
  We note that {\FEii} lines are seen on its optical maximum \citep{ATel_14577,ATel_14614}, probably following the temperature decrease.
}.

Notably, the strong lines of {\Niii} (4638 {\AA}) and {\HEii} (4686 {\AA} and 5412 {\AA}) were detected only at $t$ = +9.88 hours (Figure \ref{fig:v1405cas-spec}b, c).
The spectrum, in which the {\Niii} triplet (4638 {\AA}) is detected but {\Niv} lines (e.g., 4058 {\AA}, 7109--7123 {\AA}) are not, is similar to those of late O-type stars \citep{Gray_Corbally_2009}, which are characterized by high-ionization lines due to high temperature.
Also, the {\Ciii} line (5696 {\AA}), overlapping with the {\Nii} line (5679 {\AA}), is detected only at $t$ = +9.88 hours (Figure \ref{fig:v1405cas-spec}c).
Between $t$ = +9.88 hours and $t$ = +23.77 hours, the {\HEii}, {\Niii}, and {\Ciii} lines almost disappeared while the {\Nii} line (5679 {\AA}) became stronger (Figure \ref{fig:v1405cas-spec}b, c)\footnote{
  From the HDS spectrum (\S\ref{subsec:spectra:species_HDS:4600}), we consider that the complex between 4600 {\AA} and 4700 {\AA} observed at $t$ = +23.77, +33.94, +71.79, and +81.90 hours (Figure \ref{fig:v1405cas-spec}b) can be explained by other lines, such as {\Nii} (4631/4643 {\AA}) or {\ALii} (4663{\AA}).\label{foot:bowen}
}.
Between $t$ = +33.94 and $t$ = +71.79 hours, the {\Nii} line was weakened, and the {\SIii} (5958/5979 and 6347/6371 {\AA}), {\Oi} (7773 {\AA}), and the {\ALii} (6237 {\AA}) lines became stronger (Figure \ref{fig:v1405cas-spec}c, d, e).
The {\Nii}, {\ALii}, {\SIii}, and {\ALiii} (5696.604/5722.730 {\AA}) show multiplets as identified in the HDS spectrum (Figure \ref{fig:v1405cas-spec-HDS}; \S\ref{subsec:spectra:species_HDS}).

We extracted the line luminosities from our spectra; the extinction was corrected for assuming $E(B - V)$ = 0.51 \citep{ATel_14476} and $R_V$ = 3.1, and then the flux was converted to the luminosity scale, adopting the Gaia distance (\S\ref{sec:v1405cas}).
The line luminosities at different epochs are summarized in Table \ref{tab:line_luminosity}.

The {\Niii} and {\HEii} lines are rarely found in the early spectra of novae showing the rapid rise.  
Indeed, our spectrum may be a less extreme version of the initial spectrum taken for the recurrent nova T Pyx in 2011 \citep{Arai_2015} at 4.4 hours since the discovery; it showed not only {\HEii} and {\Niii} lines but also {\Civ} and {\Niv} lines.
It is likely that these high-ionization lines are strong only in the short time scale just after the nova eruption;
such very early-phase spectra are largely missing in the literature.
Another important object in this context is the extremely slow nova Gaia22alz \citep{Aydi_2023,ATel_15270}; its first spectrum, which was taken when it was only $\lesssim$ 4 magnitude above its quiescent, resembled that of T Pyx in the existence of prominent Balmer, {\HEi}, {\HEii}, {\Niii}, and {\Civ} lines.

\subsection{Line Identifications with the HDS Spectrum}
\label{subsec:spectra:species_HDS}
The HDS spectrum, with its high resolution and signal-to-noise ratio, allows to separately detect each component of some multiplets.
The spectrum thus helps robust identifications of lines when several possibilities are not resolved by the low-resolution data.
In this section, we provide line identifications using the HDS spectrum.

\subsubsection{4600 {\AA} -- 4700 {\AA}}
\label{subsec:spectra:species_HDS:4600}
The local maxima between 4600 and 4650 {\AA} in the HDS spectrum are consistent with the {\Nii} sextuplet (4601.478/4607.153/4613.868/4621.393/4630.539/4643.086 {\AA}; Figure \ref{fig:v1405cas-spec-HDS}a); on $t$ = +53.53 h, the {\Niii} (4638 {\AA}) and {\Ciii} (4650 {\AA}) lines have already weakened.
The {\ALii} singlet (4663.046 {\AA}) is also detected, corroborating our identification of the {\ALii} (6237 {\AA}) line (\S\ref{subsec:spectra:species}, \ref{subsec:spectra:species_HDS:6200}).

\subsubsection{5660 {\AA} -- 5730 {\AA}}
\label{subsec:spectra:species_HDS:5700}
We detected four of the {\Nii} sextuplet (with the components at 5666.629/5679.558/5686.213/5710.766 {\AA} detected; those at 5676.017/5730.656 {\AA} non-detected), as well as the {\ALiii} doublet (5696.604/5722.730 {\AA}) in the HDS spectrum (Figure \ref{fig:v1405cas-spec-HDS}b).
The presence of the partner ({\ALiii} 5722.730 {\AA}) and the consistency in the relative line strengths (\S\ref{sec:1zone}) justify the identification of the line around 5697 {\AA} in the HDS spectrum ($t$ = +53.53 hours) as {\ALiii} (5696.604 {\AA}) rather than the {\Ciii} singlet (5695.916 {\AA}).
On the other hand, the component there on $t$ = +9.88 hours, as seen in the low-resolution KOOLS-IFU spectrum, is more likely {\Ciii} rather than {\ALiii} (Figure \ref{fig:v1405cas-spec}c; \S\ref{sec:1zone}).

\subsubsection{6200 {\AA} -- 6400 {\AA}}
\label{subsec:spectra:species_HDS:6200}
The line frequently refereed to as `{\ALii} 6237 {\AA}' \citep{Williams_1992} is actually a triplet (6226.207/6231.7/6243.2 {\AA}; more strictly, it is a sextuplet while some of them are not resolved even in the high-dispersion spectrum).
While the triplet is seen as a single line in the KOOLS-IFU spectra (`{\ALii} 6237 {\AA}'; Figure \ref{fig:v1405cas-spec}), it is resolved into the fine structure in the high-resolution HDS spectrum (Figure \ref{fig:v1405cas-spec-HDS}c), showing cleaer detection of the 6231.7 and 6243.2 {\AA} components. 
The wavelengths and relative line strengths (including non-detection of the 6226.207 {\AA} component) are consistent with the atomic data of the triplet, which justifies our identification.

\subsection{Evolution of the Line Profiles}
\label{subsec:spectra:prifiles}
The H$\beta$ and the {\HEi} (7065 {\AA}) lines show the P Cygni profile (Figures \ref{fig:v1405cas-spec} and \ref{fig:v1405cas-spec-vel}), similar to most of the lines that novae typically show around their maxima.
The velocity of the blue-shifted absorption minimum of the H$\beta$ line shifts redward from $-$1.6(1)$\times$10$^{\text{3}}$ km s$^{-\text{1}}$ on $t$ = +9.88 hours to $-$1.3(7)$\times$10$^{\text{3}}$ km s$^{-\text{1}}$ on $t$ = +81.90 hours. 
Similarly, the {\HEi} (7065 {\AA}) line seen in the VPH-red spectra shows a decreasing velocity from $-$1.6(6)$\times$10$^{\text{3}}$ km s$^{-\text{1}}$ on $t$ = +9.88 hours to $-$1.3(3)$\times$10$^{\text{3}}$ km s$^{-\text{1}}$ on $t$ = +81.90 hours.
Such decreases in the velocities of the absorption components are also found in the H$\alpha$ and {\HEi} (4471 {\AA}) lines.
While the velocity measured here using the KOOLS-IFU spectra involves an error of several percent as seen in the comparison between the original HDS spectrum and the one for which the resolution is decreased to the level of the KOOLS-IFU spectra (Figure \ref{fig:v1405cas-spec-vel}a), the deceleration trend is robustly seen beyond the error level.

The deceleration seen in the absorption velocities indicates a homogeneously expanding, ejecta-like outflow, rather than (quasi-)steady wind-like outflow, for the site of the spectral formation in the very early phase \citep[e.g.,][]{Aydi_2020}.
In the ejecta-like outflow, the velocity is usually faster at outer regions (e.g., the Hubble flow).
The photosphere is expected to recede in the mass coordinate while it expands outward in radius; this behavior could lead to the continuous decrease both in the velocity and temperature, as a general expectation.
In contrast, in case of the steady-state wind-like outflow, the velocity and density profiles do not necessarily evolve quickly.

\section{Analyses of the Line Luminosities}
\label{sec:1zone}

\subsection{Method}
\label{subsec:1zone:method}
To diagnose the physical conditions responsible for the spectral line formation, we perform crude analyses of the line strengths under the one-zone approximation.
The method of the analyses here is briefly described below (for detail, see Appendix).
There are four input parameters; the temperature ($T$), the gas density ($\rho$), the composition, and the volume $V$.
For each element, the ionization fractions and level populations are determined by assuming the Local-Thermodynamic Equilibrium (LTE) condition, which are then used to compute the line emissivities, through spontaneous emission and recombination.
The line luminosities are obtained by multiplying the line emissivities by $V$.

In addition to the line luminosities, we also estimate the optical depth of the system to the electron scattering; $\tau_{\rm e} = n_{\rm e}\sigma_{\rm e}R$ with $V = \frac{\text{4}\pi}{\text{3}}R^{\text{3}}$, where $n_{\rm e}$ is the electron number density, $\sigma_{\rm e}$ is the Thomson cross section, and $R$ is the radius.
In order to reduce the number of unknown parameters, we further replace the radius and volume of the system with the velocity and time since the outburst (which are both observables), assuming homologous expansion for the dynamics; $R = vt_{\rm ej}$, with $v$ = 2000 km s$^{-{\text{1}}}$ and $t_{\rm ej}$ = 1, 2, and 4 days for $t$ = +9.88, +33.94, and +81.90 hours.
The other three parameters ($\rho$, $T$, and the composition) are allowed to vary.
Note that these three parameters are determined by the line strength ratios through the local conditions, and do not depend on the assumed expansion dynamics.
Only the conversion from the density to the mass is affected by the size and the volume of the system.

\subsection{Line Strength Diagnostics}
\label{subsec:1zone:result}

\begin{figure*}[htb]
  \centering
  \includegraphics[width=\hsize]{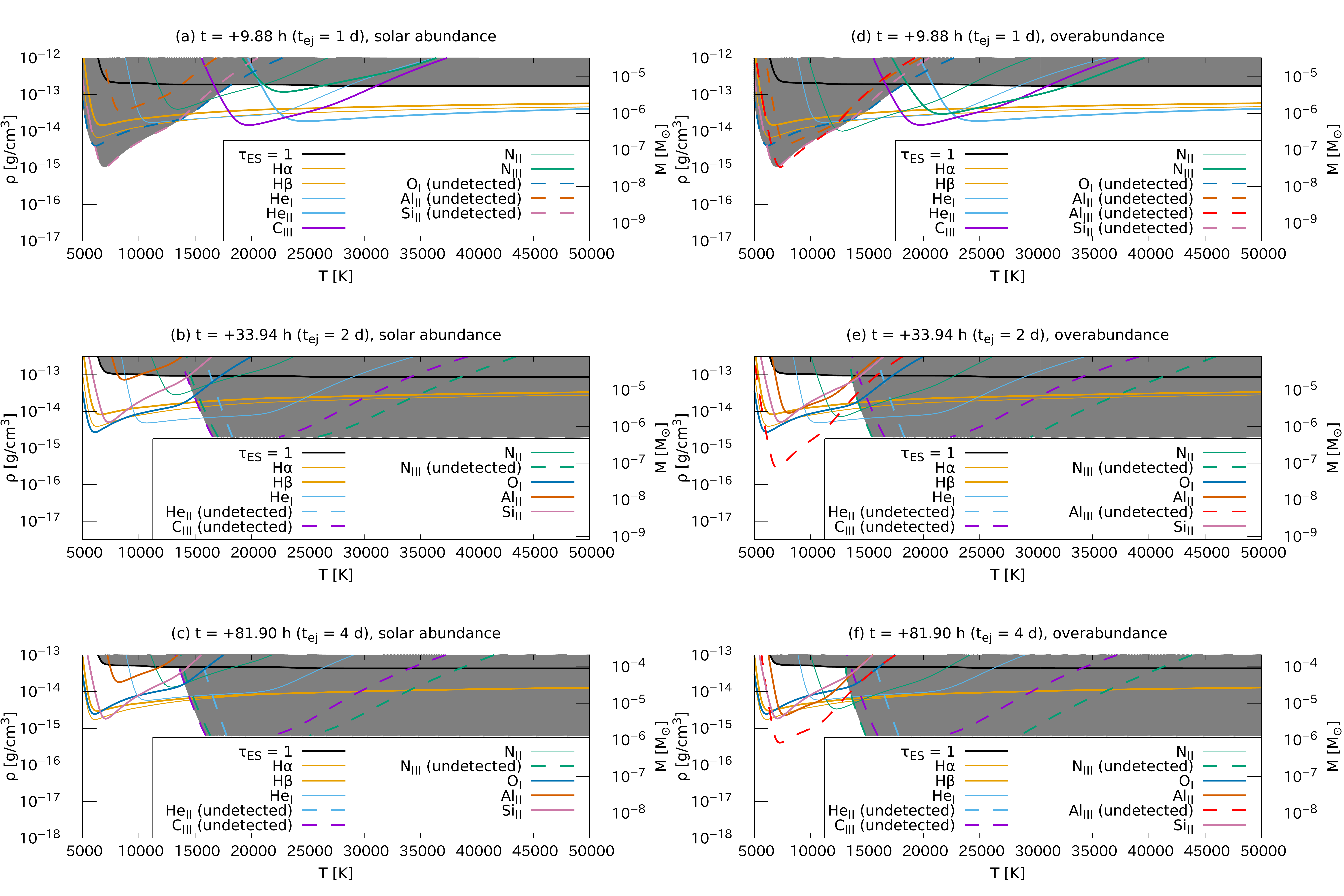}
  \caption{
    Constraints on the conditions of the line-forming region in the $(T, \rho)$ plane for (a,d) $t$ = +9.88, (b,e) $t$ = +33.94, and (c,f) $t$ = +81.90 hours.
    The observed line luminosities are reproduced by the one-zone model on the colored curves, where different colors are used for different lines; a consistent solution is therefore found for the position of ($T$, $\rho$) where all the lines are largely overlapping with each other. 
    For the non-detected lines, we place the limits in the ($T$, $\rho$) plane as shown by the colored dashed lines (with different colors for different lines), using the typical sensitivity of  $\text{10}^{\text{32}}$ erg s$^{-1}$ for the non-detection. 
    The black line in each panel shows the constraint of the electron scattering optical depth being unity, above which the line formation will become ineffective.
    The shaded area in each panel shows the `forbidden' region, where either at least one of the non-detection constraints or electron-scattering optical depth constraint is not satisfied. 
    The left panels adopt the solar abundance for all the elements, while the right panels adopt the enhanced mass fractions for nitrogen and aluminum, by a factor of {\exN} and {\exAL} relative to the solar abundance, respectively.
  }
  \label{fig:line_luminosity}
\end{figure*}

The results of our calculations are compared with the observations in Figure \ref{fig:line_luminosity}.
Each solid line shows the constraint on a combination of ($T$, $\rho$) through the observed luminosity of each line, i.e., the model provides the correct luminosity if $T$ and $\rho$ are on the line.
The exercise is done for the data at three epochs (+9.88, +33.94, and +81.90 hours, from top to bottom in Figure \ref{fig:line_luminosity}).
The left panels show the result adopting the solar abundance \citep{Allens_astrophysical_quantities}, while the right panels show the model results adopting the enhancement of nitrogen and aluminum by a factor of {\exN} and {\exAL} relative to the solar abundance. 

On the same plots, we also show the constraint from the non-detection of some lines.
The dashed lines show combinations of ($T$, $\rho$) that predict line luminosities of 10$^{\text{32}}$ erg s$^{-{\text{1}}}$ for some emission lines that are not detected.
The adopted luminosity here is a typical upper limit for the non-detection; the region in the ($T$, $\rho$) plane above at least one of these dashed lines is thus regarded as `forbidden'.
In addition to this constraint from the undetected lines, we also use the electron scattering optical depth as another constraint; we regard the region where the optical depth exceeds unity as being `forbidden', as spectral lines will likely be killed once the system suffers from repeated multiple scatterings.
The `forbidden' region is described by the shaded area in Figure \ref{fig:line_luminosity}.
Assuming the homologously expanding ejecta, the density is converted to the mass ($M = \rho V$) as shown in the right vertical axes.

Panel (a) shows that the line luminosities of the H$\alpha$, H$\beta$, {\HEi}, {\HEii}, and {\Ciii} are all explained by $T \sim$ 22500 K and $\rho \sim$ 3 $\times$ 10$^{-\text{14}}$ g cm$^{-\text{3}}$ at $t$ = +9.88 hours, with the solar abundance\footnote{
  For the density and temperature considered here, the strength of {\Ciii} 4650 {\AA} is negligibly small as compared to {\Niii} 4640 {\AA}.
  \label{foot:bowen_Ciii}
}.
This condition does not conflict with the non-detection constraints as well.
The ejecta mass corresponding to the density here is $\sim$ 10$^{-\text{6}}$ $M_{\odot}$.
However, the high luminosities of the nitrogen lines at $t$ = +9.88 hours are not explained by the solar abundance composition.
Panel (d) shows that these line strengths are roughly explained if the mass fraction of nitrogen is enhanced by a factor of {\exN} relative to the solar abundance.

Similar comparisons between the observed and the model line luminosities are shown in panels (b) and (c) for $t$ = +33.94 and +81.90 hours, respectively.
On $t$ = +33.94 hours, the line luminosities of the H$\alpha$, H$\beta$, {\HEi}, {\Oi}, and {\SIii} are consistently explained by $T$ $\sim$ 10000 K and $\rho \sim$ 10$^{-{\text{14}}}$ g cm$^{-{\text{3}}}$, assuming the solar abundance.
Similarly, $T$ $\sim$ 10000 K and $\rho \sim$ 3 $\times$ 10$^{-{\text{15}}}$ g cm$^{-{\text{3}}}$ explain the luminosities on $t$ = +81.90 hours.
Again, the model under-reproduces the {\Nii} luminosity; this is remedied by increasing the nitrogen enhancement by a factor of {\exN} as shown in panels (e) and (f), as is consistent with the same enhancement adopted for the earlier epoch. 

The strong {\ALii} appears in the spectrum on +33.94 hours, and the similar situation with the nitrogen lines is found for the {\ALii} line; the solar abundance does not produce the {\ALii} luminosity as high as observed, in the physical condition required for the other lines.
As shown in panel (e), if the mass fraction of aluminum is enhanced by a factor of $\sim$ {\exAL} relative to the solar abundance, the observed luminosity is reproduced.
We confirm that the enhancement of nitrogen and aluminum does not introduce inconsistency with the upper limit placed for the non-detected lines.
Essentially the same argument applies to the spectrum on +81.90 hours; we need the enhancement of nitrogen and aluminum by a factor of $\sim$ {\exN} and {\exAL}, respectively, while keeping the solar abundance for the other elements (panels c and f).

We find that the density and temperature of the line-forming region both decrease as time goes by.
This is most simply explained by the expanding `ejecta' or `fireball', on top of the expanding photosphere, rather than the wind-like mass outflow, as is consistent with the line-velocity evolution (\S\ref{subsec:spectra:prifiles}).
If it is interpreted as the homologously expanding ejecta, the mass of the line-forming region also increases for later epochs.
This can be explained by increasing transparency due to the expansion, again consistent with the expanding ejecta picture.

\section{Discussion}
\label{sec:discussion}

\subsection{A Low Mass ONeMg WD behind V1405 Cas}
\label{subsec:discussion:v1405cas}
We have found the enhancement of the mass fraction of aluminum in V1405 Cas, by a factor of $\sim$ {\exAL} relative to the solar abundance (\S\ref{subsec:1zone:result}), based on the detection of the {\ALii} 6237 {\AA} line.
From nucleosynthesis models of novae \citep[e.g.,][]{Politano_1995,Jose_1998,Jose_2006,Iliadis_2015,Jose_2020}, such a large excess of aluminum in a nova indicates an ONeMg WD progenitor rather than a CO WD, because the Ne-Na and Mg-Al cycles work much more efficiently in the ONeMg WD case.
Therefore, we conclude that the WD behind V1405 Cas is likely an ONeMg WD, as is consistent with the neon-nova classification reported by \cite{ATel_15796} based on its spectra in the decay phase\footnote{
  We indeed had reached this conclusion before \cite{ATel_15796} classified V1405 Cas as a neon nova (e.g., \url{https://sites.google.com/view/supervirtual2022/home}).\label{foot:supervirtual}
}.
The low-mass progenitor for V1405 Cas is also consistent with the spectral property; it shows narrower emission lines (cf., \S\ref{subsec:spectra:species}) than novae hosting massive WDs like U Sco \citep{Yamanaka_2010}.

To our knowledge, V723 Cas is the only previous example of a very slow nova classified as a neon nova.
Indeed, the observational properties of V723 Cas show similarities to V1405 Cas.
After the maximum light, V723 Cas showed humps with repeated brightening and fading in its light curve, a striking property that is very similar to what is seen in V1405 Cas.
The WD mass, $M_{\rm WD}$, is estimated to be below 0.7 $M_{\odot}$ for V723 Cas from the light curve \citep{Hachisu_2015}.
Therefore V723 Cas had been initially considered as a CO nova \citep{Hachisu_2004,Iijima_2006}.
However, \cite{Iijima_2006} derived the abundance of neon in V723 Cas as $\log(N_{\text{Ne}}/N_{\text{H}}) = -\text{2.5}$ (i.e., Ne/H $\sim$ 26 (Ne/H)${}_{\odot}$), and \cite{Takeda_2018} summarized that V723 Cas is the first example of a very slow nova showing a signature of being a neon nova, hosting a WD of only $\sim$ 0.6 $M_\odot$.

V1405 Cas is the second example of this kind; a very slow nova classified as a neon nova.
Such a low mass ONeMg WD is not expected in the standard theory of single-star evolution (\S\ref{sec:intro}).
However, as proposed by \cite{Kepler_2016} and \cite{Takeda_2018}, it may be explained by extensive mass loss through binary evolution, a late violent thermal pulse, or a C-shell flash.
In these cases, the WD has to lose $\sim$ 0.5 $M_\odot$ since the formation, and whether such an evolution is possible has not been clarified.
Further study to identify possible evolutionary channel(s) for the formation of a low-mass ONeMg WD is required, which hopefully will deepen our understanding of the formation and evolution of WDs.

\subsection{Aluminum overabundance and neon novae}
\label{subsec:discussion:past_neon_novae}
As argued in the previous section, the aluminum overabundance is expected for neon novae according to nucleosynthesis calculations.
There is indeed a hint of this signature seen in previous observations of novae in the decline phases.
For example, V838 Her, which is one of the fastest neon novae, has aluminum abundance Al/H $\sim$ 23 (Al/H)${}_{\odot}$ as was derived based on ultraviolet observations \citep{Schwarz_2007}.
Likewise, based on UV observations, the aluminum excesses have been derived for very fast neon novae LMC 1990\#1 \citep[Al/H $\sim$ 292 (Al/H)${}_{\odot}$;][]{Vanlandingham_1999}, V382 Vel \citep[Al/H $\sim$ 20.6 (Al/H)${}_{\odot}$;][]{Shore_2003}, V351 Pup \citep[Al/H $\sim$ 140 (Al/H)${}_{\odot}$;][]{Saizar_1996}, and V693 CrA \citep[Al/H $\sim$ 64 (Al/H)${}_{\odot}$;][]{Vanlandingham_1997}, and fast neon novae V1974 Cyg \citep[Al/H = 8.2 (Al/H)${}_{\odot}$;][]{Austin_1996} and QU Vul \citep[Al/H $\sim$ 39-376 (Al/H)${}_{\odot}$;][]{Saizar_1992,Andrea_1994,Schwarz_2002}.
This is further added by the aluminum excess derived for neon novae based on NIR observations; an example is the very fast nova V1500 Cyg (Al/H $\geq$ 22 (Al/H)${}_{\odot}$).
These examples hint the possible relation between the aluminum excess and neon novae (i.e., ONeMg WD progenitors), as expected from the nucleosynthesis point of view.
However, we emphasize that not all novae showing aluminum excesses are neon novae; \cite{Orio_2020} reported Al/Al${}_{\odot}$ = $\text{172}_{-\text{63}}^{+\text{153}} - \text{0.0}^{+\text{1.0}}$ from the \textit{Chandra} observation of the symbiotic recurrent CO nova V3890 Sgr.

Indeed, it is further complicated by the case of V723 Cas --  while it was reported as the first case of the `very slow neon nova' (see above), the aluminum abundance is not enhanced; the photoionization model \citep{Takeda_2018} for the data taken by Keck-OSIRIS \citep{Lyke_2009} indicates Al/H = 1.3 (Al/H)${}_{\odot}$, unlike the case for V1405 Cas.
Therefore, the relation between the aluminum enhancement and the ONeMg WD progenitor requires further investigation and clarification.

\subsection{{\ALii} $\lambda$6237 as the earliest indicator for a neon nova}
\label{subsec:discussion:6237}
Neon novae are rare.
This is especially the case for slow {\em and} neon novae.
As such, it is useful to identify neon nova candidates as early as possible before the definitive classification is given through the late-time neon lines, so that intensive follow-up observations can be coordinated from the early phase.

The present work suggests that {\ALii} $\lambda$6237 can be a useful predictor that can be used as the earliest footprint to identify a neon nova `candidate' (with the caveat that the relation between the aluminum enhancement and the neon nova is yet to be established; \S\ref{subsec:discussion:past_neon_novae}).
As shown in the present work, V1405 Cas shows the rapid decrease in the temperature in the initial rising phase.
While the spectral observations in such short initial brightening phases are still lacking for novae (except for V1405 Cas and T Pyx), we expect that it is a general behavior of novae.
In the subsequent pre-maximum halt phase, the high density of the emitting region (as shown in the present work) allows sufficiently strong {\ALii} $\lambda$6237, while another potentially strong line, {\ALii} $\lambda$7049, is easily masked by {\HEi} (7065 {\AA}) line; it is thus expected that {\ALii} $\lambda$6237 first emerges as the earliest prophet for a neon nova in the high-temperature condition (i.e., He/N-type), which is later added by {\ALii} $\lambda$7049 once the temperature decreases sufficiently so that the He lines disappear (i.e., {\FEii}-type)\footnote{
  Indeed, {\ALii} $\lambda$7049 was identified in the {\FEii}-type spectrum of fast neon nova V2264 Oph \citep{Williams_1994}.
  In the subsequent He/N-type spectrum with [{\Oi}] (their Fig. 7), {\ALii} $\lambda\lambda$ 3901, 4863, and 10090 were identified, and potential {\ALii} $\lambda$ 6237, which they did not identify, is seen.
}.

This is especially useful for slow novae.
Fast novae are generally believed to have a low density in the ejecta (due to the small ejecta mass and the high expansion velocity), and thus the phase where {\ALii} $\lambda$6237 emerges is likely very limited, or even lacking.
On the other hand, slow novae likely have high-density ejecta especially in the earliest phase, and thus detection of {\ALii} $\lambda$6237, if one is indeed a neon nova, is expected if sufficiently early-phase spectral observations are conducted.
This has been demonstrated by the present work; it helps to organize further follow-up observations not to miss an opportunity to form intensive data set for these rare but important events.

\section{Concluding Remarks}
\label{sec:conclusion}
We have presented our spectroscopic observations of the very slow nova V1405 Cas in six epochs between $t$ = +9.88 hours and $t$ = +81.90 hours after its discovery, covering between the end of the initial brightening and the earliest stage of the pre-maximum halt.
The key findings are summarized as follows;
\begin{itemize}
\item
  V1405 Cas initially showed high-ionization lines (e.g., {\Niii}), which are rapidly replaced by lower-ionization lines (e.g., {\Nii}).
\item
  We see a rapid decrease in the density and temperature, as well as an increase in the emitting mass.
  These are consistent with the idea that the initial phase of a nova is well described as an expanding fireball on top of an expanding photosphere.
\item
  The {\ALii} line (6237 {\AA}) is detected on $t$ = +23.77, +33.94, +53.53, +71.79, and +81.90 hours.
  We have estimated that aluminum is overproduced in the nova ejecta, by a factor of $\sim$ {\exAL} relative to the solar abundance.
  Similarly, nitrogen is enhanced by a factor of $\sim$ {\exN}.
\end{itemize}

We have concluded that V1405 Cas likely hosts a low-mass ONeMg WD, strengthening its classification as a very slow {\em and} neon nova; this is the second example of this kind.
\begin{itemize}
\item
  V1405 Cas is similar to V723 Cas in characteristic observational features; a very slow light curve with humps seen after the maximum light, as well as the neon nova classification based on coronal neon lines in the decline phase.
  We suggest that the aluminum overabundance in V1405 Cas supports its origin as a low-mass ONeMg WD.
\item
  Motivated by the results of reaction-network model calculations that generally predict a huge enhancement of aluminum abundance in a neon nova as compared to a CO nova, together with the aluminum enhancement seen in V1405 Cas, we have investigated a possible relation between the aluminum enhancement and their classification.
  While the survey is neither complete nor unbiased, there is a hint that fast novae showing aluminum excesses tend to be neon novae, indicating that this relation is roughly the case.
  However, the exceptions are also found (i.e., CO novae with the aluminum enhancement, as well as neon novae with the aluminum non-enhancement; the latter is the case for V723 Cas), and thus further investigation is required before establishing the relation.
\end{itemize}

Further theoretical studies are necessary to investigate the properties of these very slow novae; neon novae having such a low-mass ONeMg WD progenitor have not been investigated by model calculations.

Novae have been considered to be an important source of some isotopes, such as $^{\text{7}}$Li, $^{\text{13}}$C, $^{\text{15}}$N, $^{\text{17}}$O, $^{\text{22}}$Na, and $^{\text{26}}$Al \citep{Gerhz_1998}.
Recently, explosive $^{\text{7}}$Li production in novae has been observationally established \citep[e.g.,][]{Tajitsu_2015}.
$^{\text{26}}$Al is famous for the 1.8 MeV nuclear $\gamma$-rays, which are observed along the Galactic plane and therefore considered to originate from star-forming regions (i.e., massive stars and supernovae) \citep{Mahoney_1982,Mahoney_1984,Diehl_1995,Diehl_2006}.
However, it has been reported that some grains have isotopic ratios of $^{\text{26}}$Al/$^{\text{27}}$Al consistent with (ONe) nova models \citep{Amari_2001} and that 1/3 of $^{\text{26}}$Al in LMC is predicted to be produced by novae according to a recent chemical evolution model \citep{Vasini_2023}.
These examples show potentially important roles played in general by novae, and some unique roles by neon novae, in the origin of elements and isotopes.

Given the proximity of V1405 Cas that allows further monitoring observations even toward its staying back to the quiescent state, further follow-up observations are encouraged to deepen our understanding of neon novae and their contribution to the Galactic chemical evolution.
In addition, with {\ALii} $\lambda$6237 (and later $\lambda$7049) as the earliest indicator for a neon nova, we hope to substantially expand the sample of neon novae with intensive observational data already from the infant phase.

\begin{acknowledgments}
  We appreciate the referee for reading our manuscript carefully and giving us many constructive comments, which have helped improve the quality of the paper into a clearer and more balanced one.
  This study makes use of data obtained by the 3.8-m Seimei telescope through the programs, 21A-N-CT06 in the open-use of the observing time provided by NAOJ and 21A-K-0017 in the Kyoto University time.
  This research is based in part on data collected at the Subaru Telescope, which is operated by the National Astronomical Observatory of Japan. We are honored and grateful for the opportunity of observing the Universe from Maunakea, which has the cultural, historical and natural significance in Hawaii. 
  We acknowledge the AAVSO International Database contributed by observers worldwide.
  This work has made use of data from the European Space Agency (ESA) mission {\it Gaia} (\url{https://www.cosmos.esa.int/gaia}), processed by the {\it Gaia} Data Processing and Analysis Consortium (DPAC, \url{https://www.cosmos.esa.int/web/gaia/dpac/consortium}). Funding for the DPAC has been provided by national institutions, in particular the institutions participating in the {\it Gaia} Multilateral Agreement.
  This work is supported by JST SPRING grant JPMJSP2110 (Kenta Taguchi) and by JSPS KAKENHI grants JP18H05223, JP20H04737, JP20H00174 (Keiichi Maeda), JP22K03676 (Masayuki Yamanaka), JP20K14521, JP18H05439 (Keisuke Isogai), 21J22351 (Yusuke Tampo), and JP21K03616 (Taichi Kato).
\end{acknowledgments}

\facility{AAVSO}

\begin{deluxetable*}{ccccccccccccccc}
  \tablecaption{Observed line luminosities.}
  \label{tab:line_luminosity}
  \tabletypesize{\scriptsize}
  \tablehead{
    \colhead{$t$ [hours]} & \colhead{$t_{\rm ej}$ [days]\ref{foot:tab_tej}} & \colhead{H$\alpha$} & \colhead{H$\beta$} & \colhead{{\HEi} $\lambda$5876} & \colhead{{\HEii} $\lambda$4686} & \colhead{{\Ciii} $\lambda$5696} & \colhead{{\Nii} $\lambda$5679} & \colhead{{\Niii} $\lambda$4640} & \colhead{{\Oi} $\lambda$7773} & \colhead{{\ALii} $\lambda$6237} & \colhead{{\SIii} $\lambda\lambda$6347/6371}
  }
  \startdata
  +9.88 & 1 & 17 & 11 & 2 & 6 & 0.14 & 0.48 & 8${}^{\text{\ref{foot:bowen_Ciii}}}$ & -- & -- & --
  \\
  +33.94 & 2 & 50 & 30 & 14 & --${}^{\text{\ref{foot:bowen}}}$ & -- & 2.2 & --${}^{\text{\ref{foot:bowen}}}$ & 0.04 & 0.3 & 1
  \\
  +81.90 & 4 & 50 & 20 & 9 & --${}^{\text{\ref{foot:bowen}}}$ & -- & 2.8 & --${}^{\text{\ref{foot:bowen}}}$ & 0.15 & 0.12 & 0.9
  \enddata
  \tablecomments{
    Line luminosities are measured using SPLOT task of IRAF and tabulated in units of $\text{10}^{\text{34}}$ erg s$^{-{\text{1}}}$.
    \begin{enumerate}[label={$^{\text{\alph{enumi}}}$}]
    \item\label{foot:tab_tej}
      See \S\ref{subsec:1zone:method} and Appendix.
    \end{enumerate}
  }
\end{deluxetable*}

\appendix
\section{Computations of the Line Strengths}
We approximate the line luminosity of a transition from upper level $u$ to lower level $l$ as the sum of the following two components,
\begin{equation}
  L_{u \to l} \sim L_{u \to l}^{\text{spon}} + L_{u \to l}^{\text{recomb}},
\end{equation}
where $L_{u \to l}^{\text{spon}}$ is the spontaneous component for given level populations, and $L_{u \to l}^{\text{recomb}}$ is the recombination component for given ionization status.
The spontaneous component is expressed as
\begin{equation}
  L_{u \to l}^{\text{spon}} = \frac{h\nu_{ul}}{4\pi} n_u A_{ul} V ,
\end{equation}
where $\nu_{ul}$ is the line frequency, $n_u$ is the number density of the upper level, $A_{ul}$ is Einstein's A coefficient, and $V$ is the volume of the system (see below). 
The recombination component can be evaluated as
\begin{equation}
  L_{u \to l}^{\text{recomb}} = \frac{h\nu_{ul}}{4\pi} n_{\text{ion}} n_{\rm e} \alpha^{\text{(total)}}_{\text{ion} \to u} (T) P^{(\text{direct})}_{u \to l} V,
\end{equation}
where $n_{\text{ion}}$ is the number density of the ion under consideration\footnote{
  For simplicity, only the ground state is considered here.
} that leads to recombination cascade after recombination with a free electron, $n_{\rm e}$ is the number density of electrons, $\alpha^{\text{(total)}}_{\text{ion} \to u}(T)$ is the `total' recombination coefficient (defined later) to level $u$, and $P^{\text{(direct)}}_{u \to l}$ is the probability that an ion at level $u$ transitions to level $l$ `directly' (i.e., without transitioning to intermediate states before going down to level $l$).

$P^{\text{(direct)}}_{u \to l}$ is determined by the Einstein's A coefficients as follows; 
\begin{equation}
  P^{\text{(direct)}}_{u \to l} = \frac{A_{ul}}{\sum_{l' < u} A_{ul'}},
\end{equation}
where $\sum_{l' < u}$ means that it is summed over all the levels $l'$ whose energy eigenvalues are lower than that of level $u$.

The `total' recombination coefficient $\alpha^{\text{(total)}}_{\text{ion} \to u}(T)$ is defined as
\begin{equation}
  \alpha^{\text{(total)}}_{\text{ion} \to u}(T) = \alpha^{\text{(direct)}}_{\text{ion} \to u} (T) + \sum_{u' > u}\alpha^{\text{(direct)}}_{\text{ion} \to u'} (T)P^{\text{(total)}}_{u' \to u},
\end{equation}
where
\begin{equation}
  P^{\text{(total)}}_{u \to l} = P^{\text{(direct)}}_{u \to l} + \sum_{l < m < u} P^{\text{(direct)}}_{u \to m} P^{\text{(total)}}_{m \to l}
\end{equation}
is the `total' probability that an ion at level $u$ transitions to level $l$ through any combinations of different transitions, and $\alpha^{\text{(direct)}}_{\text{ion} \to u}(T)$ is the `direct' recombination coefficients to level $u$.
The recombination coefficients can be evaluated by the photoionization cross sections since the relation must be applicable to the thermal-equilibrium limit\footnote{
  For simplicity, we focus only on spontaneous radiative recombination and ignore the effect of induced radiative recombination.
}.
Therefore, $\alpha^{\text{(direct)}}_{\text{ion} \to u}(T)$ can be written as
\begin{equation}
  \alpha^{\text{(direct)}}_{\text{ion} \to u}(T) = \frac{n_u}{n_{\text{ion}} n_{\rm e}} \int_{\nu_{\text{th}}}^{+\infty} \frac{4\pi \tilde{B}_{\nu}(T)}{h\nu}\sigma_{u \to \text{ion}}(\nu)\,{\rm d}\nu,
\end{equation}
where $\sigma_{u \to \text{ion}}(\nu)$ is the photoionization cross-section from level $u$ and
\begin{equation}
  \tilde{B}_\nu (T) = \frac{2h\nu^3}{c^2}\exp\left(-\frac{h\nu}{kT}\right).
\end{equation}

Under the one-zone approximation, the line emissivities (per volume) can be computed following the above formalism, for given level populations and ionization status (and $n_{\rm e}$), and chemical composition, for the input parameters ($\rho, T$).
In doing this, we make use of the atomic database of \textit{CMFGEN}\footnote{
  \url{https://kookaburra.phyast.pitt.edu/hillier/web/CMFGEN.htm}
} \citep{Hillier_1998,Hillier_1990}.

In the present work (\S\ref{sec:1zone}), we assume the LTE to derive level populations and ionization status (and thus $n_{\rm e}$).
Additionally, the volume must be specified to convert the line emissivities to the line luminosities.
This is then used to compute the mass of the emitting region as a result of the fit to the observed line luminosities.

In the present work, we assume the ejecta-like outflow, which is justified by the decrease in the line velocities with time (\S\ref{subsec:spectra:prifiles}).
$V$ is then related to the typical size of the system $R$ as
\begin{equation}
  V = \frac{4\pi}{3} R^3,
\end{equation}
and $R$ can be expressed as
\begin{equation}
  R \sim 2.5 \times 10^2 R_\odot \left(\frac{v}{\text{2000 km/s}}\right)\left(\frac{t_{\text{ej}}}{\text{1 day}}\right),
\end{equation}
where $v$ is the velocity of the ejecta and $t_{\text{ej}}$ is the time that has passed since the outburst.
In this paper, we adopt $v$ = 2000 km/s.
As is in table \ref{tab:line_luminosity}, we adopt $t_{\rm ej}$ = 1, 2, and 4 days for $t$ = +9.88, +33.94, and +81.90 hours, respectively.

\bibliographystyle{aasjournal}
\bibliography{v1405cas}{}

\begin{thebibliography}{}
\expandafter\ifx\csname natexlab\endcsname\relax\def\natexlab#1{#1}\fi
\providecommand{\url}[1]{\href{#1}{#1}}
\providecommand{\dodoi}[1]{doi:~\href{http://doi.org/#1}{\nolinkurl{#1}}}
\providecommand{\doeprint}[1]{\href{http://ascl.net/#1}{\nolinkurl{http://ascl.net/#1}}}
\providecommand{\doarXiv}[1]{\href{https://arxiv.org/abs/#1}{\nolinkurl{https://arxiv.org/abs/#1}}}

\bibitem[{{Amari} {et~al.}(2001){Amari}, {Gao}, {Nittler}, {Zinner},
  {Jos{\'e}}, {Hernanz}, \& {Lewis}}]{Amari_2001}
{Amari}, S., {Gao}, X., {Nittler}, L.~R., {et~al.} 2001, \apj, 551, 1065,
  \dodoi{10.1086/320235}

\bibitem[{{Andrea} {et~al.}(1994){Andrea}, {Drechsel}, \&
  {Starrfield}}]{Andrea_1994}
{Andrea}, J., {Drechsel}, H., \& {Starrfield}, S. 1994, \aap, 291, 869

\bibitem[{{Arai} {et~al.}(2015){Arai}, {Isogai}, {Yamanaka}, {Akitaya}, \&
  {Uemura}}]{Arai_2015}
{Arai}, A., {Isogai}, M., {Yamanaka}, M., {Akitaya}, H., \& {Uemura}, M. 2015,
  Acta Polytechnica CTU Proceedings, 2, 257

\bibitem[{{Austin} {et~al.}(1996){Austin}, {Wagner}, {Starrfield}, {Shore},
  {Sonneborn}, \& {Bertram}}]{Austin_1996}
{Austin}, S.~J., {Wagner}, R.~M., {Starrfield}, S., {et~al.} 1996, \aj, 111,
  869, \dodoi{10.1086/117835}

\bibitem[{{Aydi} {et~al.}(2020){Aydi}, {Chomiuk}, {Izzo}, {Harvey},
  {Leahy-McGregor}, {Strader}, {Buckley}, {Sokolovsky}, {Kawash}, {Kochanek},
  {Linford}, {Metzger}, {Mukai}, {Orio}, {Shappee}, {Shishkovsky}, {Steinberg},
  {Swihart}, {Sokoloski}, {Walter}, \& {Woudt}}]{Aydi_2020}
{Aydi}, E., {Chomiuk}, L., {Izzo}, L., {et~al.} 2020, \apj, 905, 62,
  \dodoi{10.3847/1538-4357/abc3bb}

\bibitem[{{Aydi} {et~al.}(2023){Aydi}, {Chomiuk}, {Miko{\l}ajewska}, {Brink},
  {Metzger}, {Strader}, {Buckley}, {Harvey}, {Holoien}, {Izzo}, {Kawash},
  {Linford}, {Molaro}, {Molina}, {Mr{\'o}z}, {Mukai}, {Orio}, {Panurach},
  {Senchyna}, {Shappee}, {Shen}, {Sokoloski}, {Sokolovsky}, {Urquhart}, \&
  {Williams}}]{Aydi_2023}
{Aydi}, E., {Chomiuk}, L., {Miko{\l}ajewska}, J., {et~al.} 2023, \mnras, 524,
  1946, \dodoi{10.1093/mnras/stad1914}

\bibitem[{{Barden} \& {Armandroff}(1995)}]{hydra2}
{Barden}, S.~C., \& {Armandroff}, T. 1995, in Society of Photo-Optical
  Instrumentation Engineers (SPIE) Conference Series, Vol. 2476, Fiber Optics
  in Astronomical Applications, ed. S.~C. {Barden}, 56--67,
  \dodoi{10.1117/12.211839}

\bibitem[{{Barden} {et~al.}(1994){Barden}, {Armandroff}, {Muller}, {Rudeen},
  {Lewis}, \& {Groves}}]{hydra1}
{Barden}, S.~C., {Armandroff}, T., {Muller}, G., {et~al.} 1994, in Society of
  Photo-Optical Instrumentation Engineers (SPIE) Conference Series, Vol. 2198,
  Instrumentation in Astronomy VIII, ed. D.~L. {Crawford} \& E.~R. {Craine},
  87--97, \dodoi{10.1117/12.176816}

\bibitem[{{Bode} \& {Evans}(2008)}]{Bode_Evans_2008}
{Bode}, M.~F., \& {Evans}, A. 2008, {Classical Novae}, Vol.~43

\bibitem[{{Brink} {et~al.}(2022){Brink}, {Aydi}, {Buckley}, {Chomiuk},
  {Kawash}, {Orio}, {Mikolajewska}, {Sokolovsky}, {Strader}, {Stanek},
  {Kochanek}, {Smith}, \& {Shappee}}]{ATel_15270}
{Brink}, J., {Aydi}, E., {Buckley}, D.~A.~H., {et~al.} 2022, The Astronomer's
  Telegram, 15270, 1

\bibitem[{{Cox}(2000)}]{Allens_astrophysical_quantities}
{Cox}, A.~N. 2000, {Allen's astrophysical quantities}

\bibitem[{{Diehl} {et~al.}(1995){Diehl}, {Dupraz}, {Bennett}, {Bloemen},
  {Hermsen}, {Knoedlseder}, {Lichti}, {Morris}, {Ryan}, {Schoenfelder},
  {Steinle}, {Strong}, {Swanenburg}, {Varendorff}, \& {Winkler}}]{Diehl_1995}
{Diehl}, R., {Dupraz}, C., {Bennett}, K., {et~al.} 1995, \aap, 298, 445

\bibitem[{{Diehl} {et~al.}(2006){Diehl}, {Halloin}, {Kretschmer}, {Lichti},
  {Sch{\"o}nfelder}, {Strong}, {von Kienlin}, {Wang}, {Jean}, {Kn{\"o}dlseder},
  {Roques}, {Weidenspointner}, {Schanne}, {Hartmann}, {Winkler}, \&
  {Wunderer}}]{Diehl_2006}
{Diehl}, R., {Halloin}, H., {Kretschmer}, K., {et~al.} 2006, \nat, 439, 45,
  \dodoi{10.1038/nature04364}

\bibitem[{{Flewelling} {et~al.}(2020){Flewelling}, {Magnier}, {Chambers},
  {Heasley}, {Holmberg}, {Huber}, {Sweeney}, {Waters}, {Calamida}, {Casertano},
  {Chen}, {Farrow}, {Hasinger}, {Henderson}, {Long}, {Metcalfe}, {Narayan},
  {Nieto-Santisteban}, {Norberg}, {Rest}, {Saglia}, {Szalay}, {Thakar},
  {Tonry}, {Valenti}, {Werner}, {White}, {Denneau}, {Draper}, {Hodapp},
  {Jedicke}, {Kaiser}, {Kudritzki}, {Price}, {Wainscoat}, {Chastel}, {McLean},
  {Postman}, \& {Shiao}}]{2020ApJS..251....7F}
{Flewelling}, H.~A., {Magnier}, E.~A., {Chambers}, K.~C., {et~al.} 2020, \apjs,
  251, 7, \dodoi{10.3847/1538-4365/abb82d}

\bibitem[{{Gaia Collaboration} {et~al.}(2016){Gaia Collaboration}, {Prusti},
  {de Bruijne}, {Brown}, {Vallenari}, {Babusiaux}, {Bailer-Jones}, {Bastian},
  {Biermann}, {Evans}, {Eyer}, {Jansen}, {Jordi}, {Klioner}, {Lammers},
  {Lindegren}, {Luri}, {Mignard}, {Milligan}, {Panem}, {Poinsignon},
  {Pourbaix}, {Randich}, {Sarri}, {Sartoretti}, {Siddiqui}, {Soubiran},
  {Valette}, {van Leeuwen}, {Walton}, {Aerts}, {Arenou}, {Cropper}, {Drimmel},
  {H{\o}g}, {Katz}, {Lattanzi}, {O'Mullane}, {Grebel}, {Holland}, {Huc},
  {Passot}, {Bramante}, {Cacciari}, {Casta{\~n}eda}, {Chaoul}, {Cheek}, {De
  Angeli}, {Fabricius}, {Guerra}, {Hern{\'a}ndez}, {Jean-Antoine-Piccolo},
  {Masana}, {Messineo}, {Mowlavi}, {Nienartowicz}, {Ord{\'o}{\~n}ez-Blanco},
  {Panuzzo}, {Portell}, {Richards}, {Riello}, {Seabroke}, {Tanga},
  {Th{\'e}venin}, {Torra}, {Els}, {Gracia-Abril}, {Comoretto},
  {Garcia-Reinaldos}, {Lock}, {Mercier}, {Altmann}, {Andrae}, {Astraatmadja},
  {Bellas-Velidis}, {Benson}, {Berthier}, {Blomme}, {Busso}, {Carry},
  {Cellino}, {Clementini}, {Cowell}, {Creevey}, {Cuypers}, {Davidson}, {De
  Ridder}, {de Torres}, {Delchambre}, {Dell'Oro}, {Ducourant}, {Fr{\'e}mat},
  {Garc{\'\i}a-Torres}, {Gosset}, {Halbwachs}, {Hambly}, {Harrison}, {Hauser},
  {Hestroffer}, {Hodgkin}, {Huckle}, {Hutton}, {Jasniewicz}, {Jordan},
  {Kontizas}, {Korn}, {Lanzafame}, {Manteiga}, {Moitinho}, {Muinonen},
  {Osinde}, {Pancino}, {Pauwels}, {Petit}, {Recio-Blanco}, {Robin}, {Sarro},
  {Siopis}, {Smith}, {Smith}, {Sozzetti}, {Thuillot}, {van Reeven}, {Viala},
  {Abbas}, {Abreu Aramburu}, {Accart}, {Aguado}, {Allan}, {Allasia},
  {Altavilla}, {{\'A}lvarez}, {Alves}, {Anderson}, {Andrei}, {Anglada Varela},
  {Antiche}, {Antoja}, {Ant{\'o}n}, {Arcay}, {Atzei}, {Ayache}, {Bach},
  {Baker}, {Balaguer-N{\'u}{\~n}ez}, {Barache}, {Barata}, {Barbier}, {Barblan},
  {Baroni}, {Barrado y Navascu{\'e}s}, {Barros}, {Barstow}, {Becciani},
  {Bellazzini}, {Bellei}, {Bello Garc{\'\i}a}, {Belokurov}, {Bendjoya},
  {Berihuete}, {Bianchi}, {Bienaym{\'e}}, {Billebaud}, {Blagorodnova},
  {Blanco-Cuaresma}, {Boch}, {Bombrun}, {Borrachero}, {Bouquillon}, {Bourda},
  {Bouy}, {Bragaglia}, {Breddels}, {Brouillet}, {Br{\"u}semeister},
  {Bucciarelli}, {Budnik}, {Burgess}, {Burgon}, {Burlacu}, {Busonero}, {Buzzi},
  {Caffau}, {Cambras}, {Campbell}, {Cancelliere}, {Cantat-Gaudin}, {Carlucci},
  {Carrasco}, {Castellani}, {Charlot}, {Charnas}, {Charvet}, {Chassat},
  {Chiavassa}, {Clotet}, {Cocozza}, {Collins}, {Collins}, {Costigan}, {Crifo},
  {Cross}, {Crosta}, {Crowley}, {Dafonte}, {Damerdji}, {Dapergolas}, {David},
  {David}, {De Cat}, {de Felice}, {de Laverny}, {De Luise}, {De March}, {de
  Martino}, {de Souza}, {Debosscher}, {del Pozo}, {Delbo}, {Delgado},
  {Delgado}, {di Marco}, {Di Matteo}, {Diakite}, {Distefano}, {Dolding}, {Dos
  Anjos}, {Drazinos}, {Dur{\'a}n}, {Dzigan}, {Ecale}, {Edvardsson}, {Enke},
  {Erdmann}, {Escolar}, {Espina}, {Evans}, {Eynard Bontemps}, {Fabre},
  {Fabrizio}, {Faigler}, {Falc{\~a}o}, {Farr{\`a}s Casas}, {Faye}, {Federici},
  {Fedorets}, {Fern{\'a}ndez-Hern{\'a}ndez}, {Fernique}, {Fienga}, {Figueras},
  {Filippi}, {Findeisen}, {Fonti}, {Fouesneau}, {Fraile}, {Fraser}, {Fuchs},
  {Furnell}, {Gai}, {Galleti}, {Galluccio}, {Garabato}, {Garc{\'\i}a-Sedano},
  {Gar{\'e}}, {Garofalo}, {Garralda}, {Gavras}, {Gerssen}, {Geyer}, {Gilmore},
  {Girona}, {Giuffrida}, {Gomes}, {Gonz{\'a}lez-Marcos},
  {Gonz{\'a}lez-N{\'u}{\~n}ez}, {Gonz{\'a}lez-Vidal}, {Granvik}, {Guerrier},
  {Guillout}, {Guiraud}, {G{\'u}rpide}, {Guti{\'e}rrez-S{\'a}nchez}, {Guy},
  {Haigron}, {Hatzidimitriou}, {Haywood}, {Heiter}, {Helmi}, {Hobbs},
  {Hofmann}, {Holl}, {Holland}, {Hunt}, {Hypki}, {Icardi}, {Irwin}, {Jevardat
  de Fombelle}, {Jofr{\'e}}, {Jonker}, {Jorissen}, {Julbe}, {Karampelas},
  {Kochoska}, {Kohley}, {Kolenberg}, {Kontizas}, {Koposov}, {Kordopatis},
  {Koubsky}, {Kowalczyk}, {Krone-Martins}, {Kudryashova}, {Kull}, {Bachchan},
  {Lacoste-Seris}, {Lanza}, {Lavigne}, {Le Poncin-Lafitte}, {Lebreton},
  {Lebzelter}, {Leccia}, {Leclerc}, {Lecoeur-Taibi}, {Lemaitre}, {Lenhardt},
  {Leroux}, {Liao}, {Licata}, {Lindstr{\o}m}, {Lister}, {Livanou}, {Lobel},
  {L{\"o}ffler}, {L{\'o}pez}, {Lopez-Lozano}, {Lorenz}, {Loureiro},
  {MacDonald}, {Magalh{\~a}es Fernandes}, {Managau}, {Mann}, {Mantelet},
  {Marchal}, {Marchant}, {Marconi}, {Marie}, {Marinoni}, {Marrese},
  {Marschalk{\'o}}, {Marshall}, {Mart{\'\i}n-Fleitas}, {Martino}, {Mary},
  {Matijevi{\v{c}}}, {Mazeh}, {McMillan}, {Messina}, {Mestre}, {Michalik},
  {Millar}, {Miranda}, {Molina}, {Molinaro}, {Molinaro}, {Moln{\'a}r},
  {Moniez}, {Montegriffo}, {Monteiro}, {Mor}, {Mora}, {Morbidelli}, {Morel},
  {Morgenthaler}, {Morley}, {Morris}, {Mulone}, {Muraveva}, {Musella},
  {Narbonne}, {Nelemans}, {Nicastro}, {Noval}, {Ord{\'e}novic},
  {Ordieres-Mer{\'e}}, {Osborne}, {Pagani}, {Pagano}, {Pailler}, {Palacin},
  {Palaversa}, {Parsons}, {Paulsen}, {Pecoraro}, {Pedrosa}, {Pentik{\"a}inen},
  {Pereira}, {Pichon}, {Piersimoni}, {Pineau}, {Plachy}, {Plum}, {Poujoulet},
  {Pr{\v{s}}a}, {Pulone}, {Ragaini}, {Rago}, {Rambaux}, {Ramos-Lerate},
  {Ranalli}, {Rauw}, {Read}, {Regibo}, {Renk}, {Reyl{\'e}}, {Ribeiro},
  {Rimoldini}, {Ripepi}, {Riva}, {Rixon}, {Roelens}, {Romero-G{\'o}mez},
  {Rowell}, {Royer}, {Rudolph}, {Ruiz-Dern}, {Sadowski}, {Sagrist{\`a}
  Sell{\'e}s}, {Sahlmann}, {Salgado}, {Salguero}, {Sarasso}, {Savietto},
  {Schnorhk}, {Schultheis}, {Sciacca}, {Segol}, {Segovia}, {Segransan},
  {Serpell}, {Shih}, {Smareglia}, {Smart}, {Smith}, {Solano}, {Solitro},
  {Sordo}, {Soria Nieto}, {Souchay}, {Spagna}, {Spoto}, {Stampa}, {Steele},
  {Steidelm{\"u}ller}, {Stephenson}, {Stoev}, {Suess}, {S{\"u}veges}, {Surdej},
  {Szabados}, {Szegedi-Elek}, {Tapiador}, {Taris}, {Tauran}, {Taylor},
  {Teixeira}, {Terrett}, {Tingley}, {Trager}, {Turon}, {Ulla}, {Utrilla},
  {Valentini}, {van Elteren}, {Van Hemelryck}, {van Leeuwen}, {Varadi},
  {Vecchiato}, {Veljanoski}, {Via}, {Vicente}, {Vogt}, {Voss}, {Votruba},
  {Voutsinas}, {Walmsley}, {Weiler}, {Weingrill}, {Werner}, {Wevers},
  {Whitehead}, {Wyrzykowski}, {Yoldas}, {{\v{Z}}erjal}, {Zucker}, {Zurbach},
  {Zwitter}, {Alecu}, {Allen}, {Allende Prieto}, {Amorim},
  {Anglada-Escud{\'e}}, {Arsenijevic}, {Azaz}, {Balm}, {Beck}, {Bernstein},
  {Bigot}, {Bijaoui}, {Blasco}, {Bonfigli}, {Bono}, {Boudreault}, {Bressan},
  {Brown}, {Brunet}, {Bunclark}, {Buonanno}, {Butkevich}, {Carret}, {Carrion},
  {Chemin}, {Ch{\'e}reau}, {Corcione}, {Darmigny}, {de Boer}, {de Teodoro}, {de
  Zeeuw}, {Delle Luche}, {Domingues}, {Dubath}, {Fodor}, {Fr{\'e}zouls},
  {Fries}, {Fustes}, {Fyfe}, {Gallardo}, {Gallegos}, {Gardiol}, {Gebran},
  {Gomboc}, {G{\'o}mez}, {Grux}, {Gueguen}, {Heyrovsky}, {Hoar}, {Iannicola},
  {Isasi Parache}, {Janotto}, {Joliet}, {Jonckheere}, {Keil}, {Kim},
  {Klagyivik}, {Klar}, {Knude}, {Kochukhov}, {Kolka}, {Kos}, {Kutka}, {Lainey},
  {LeBouquin}, {Liu}, {Loreggia}, {Makarov}, {Marseille}, {Martayan},
  {Martinez-Rubi}, {Massart}, {Meynadier}, {Mignot}, {Munari}, {Nguyen},
  {Nordlander}, {Ocvirk}, {O'Flaherty}, {Olias Sanz}, {Ortiz}, {Osorio},
  {Oszkiewicz}, {Ouzounis}, {Palmer}, {Park}, {Pasquato}, {Peltzer}, {Peralta},
  {P{\'e}turaud}, {Pieniluoma}, {Pigozzi}, {Poels}, {Prat}, {Prod'homme},
  {Raison}, {Rebordao}, {Risquez}, {Rocca-Volmerange}, {Rosen}, {Ruiz-Fuertes},
  {Russo}, {Sembay}, {Serraller Vizcaino}, {Short}, {Siebert}, {Silva},
  {Sinachopoulos}, {Slezak}, {Soffel}, {Sosnowska}, {Strai{\v{z}}ys}, {ter
  Linden}, {Terrell}, {Theil}, {Tiede}, {Troisi}, {Tsalmantza}, {Tur},
  {Vaccari}, {Vachier}, {Valles}, {Van Hamme}, {Veltz}, {Virtanen}, {Wallut},
  {Wichmann}, {Wilkinson}, {Ziaeepour}, \& {Zschocke}}]{Gaia_2016b}
{Gaia Collaboration}, {Prusti}, T., {de Bruijne}, J.~H.~J., {et~al.} 2016,
  \aap, 595, A1, \dodoi{10.1051/0004-6361/201629272}

\bibitem[{{Gaia Collaboration} {et~al.}(2022){Gaia Collaboration}, {Vallenari},
  {Brown}, {Prusti}, {de Bruijne}, {Arenou}, {Babusiaux}, {Biermann},
  {Creevey}, {Ducourant}, {Evans}, {Eyer}, {Guerra}, {Hutton}, {Jordi},
  {Klioner}, {Lammers}, {Lindegren}, {Luri}, {Mignard}, {Panem}, {Pourbaix},
  {Randich}, {Sartoretti}, {Soubiran}, {Tanga}, {Walton}, {Bailer-Jones},
  {Bastian}, {Drimmel}, {Jansen}, {Katz}, {Lattanzi}, {van Leeuwen}, {Bakker},
  {Cacciari}, {Casta{\~n}eda}, {De Angeli}, {Fabricius}, {Fouesneau},
  {Fr{\'e}mat}, {Galluccio}, {Guerrier}, {Heiter}, {Masana}, {Messineo},
  {Mowlavi}, {Nicolas}, {Nienartowicz}, {Pailler}, {Panuzzo}, {Riclet}, {Roux},
  {Seabroke}, {Sordo{\o}rcit}, {Th{\'e}venin}, {Gracia-Abril}, {Portell},
  {Teyssier}, {Altmann}, {Andrae}, {Audard}, {Bellas-Velidis}, {Benson},
  {Berthier}, {Blomme}, {Burgess}, {Busonero}, {Busso}, {C{\'a}novas}, {Carry},
  {Cellino}, {Cheek}, {Clementini}, {Damerdji}, {Davidson}, {de Teodoro},
  {Nu{\~n}ez Campos}, {Delchambre}, {Dell'Oro}, {Esquej},
  {Fern{\'a}ndez-Hern{\'a}ndez}, {Fraile}, {Garabato}, {Garc{\'\i}a-Lario},
  {Gosset}, {Haigron}, {Halbwachs}, {Hambly}, {Harrison}, {Hern{\'a}ndez},
  {Hestroffer}, {Hodgkin}, {Holl}, {Jan{\ss}en}, {Jevardat de Fombelle},
  {Jordan}, {Krone-Martins}, {Lanzafame}, {L{\"o}ffler}, {Marchal}, {Marrese},
  {Moitinho}, {Muinonen}, {Osborne}, {Pancino}, {Pauwels}, {Recio-Blanco},
  {Reyl{\'e}}, {Riello}, {Rimoldini}, {Roegiers}, {Rybizki}, {Sarro}, {Siopis},
  {Smith}, {Sozzetti}, {Utrilla}, {van Leeuwen}, {Abbas}, {{\'A}brah{\'a}m},
  {Abreu Aramburu}, {Aerts}, {Aguado}, {Ajaj}, {Aldea-Montero}, {Altavilla},
  {{\'A}lvarez}, {Alves}, {Anders}, {Anderson}, {Anglada Varela}, {Antoja},
  {Baines}, {Baker}, {Balaguer-N{\'u}{\~n}ez}, {Balbinot}, {Balog}, {Barache},
  {Barbato}, {Barros}, {Barstow}, {Bartolom{\'e}}, {Bassilana}, {Bauchet},
  {Becciani}, {Bellazzini}, {Berihuete}, {Bernet}, {Bertone}, {Bianchi},
  {Binnenfeld}, {Blanco-Cuaresma}, {Blazere}, {Boch}, {Bombrun}, {Bossini},
  {Bouquillon}, {Bragaglia}, {Bramante}, {Breedt}, {Bressan}, {Brouillet},
  {Brugaletta}, {Bucciarelli}, {Burlacu}, {Butkevich}, {Buzzi}, {Caffau},
  {Cancelliere}, {Cantat-Gaudin}, {Carballo}, {Carlucci}, {Carnerero},
  {Carrasco}, {Casamiquela}, {Castellani}, {Castro-Ginard}, {Chaoul},
  {Charlot}, {Chemin}, {Chiaramida}, {Chiavassa}, {Chornay}, {Comoretto},
  {Contursi}, {Cooper}, {Cornez}, {Cowell}, {Crifo}, {Cropper}, {Crosta},
  {Crowley}, {Dafonte}, {Dapergolas}, {David}, {David}, {de Laverny}, {De
  Luise}, {De March}, {De Ridder}, {de Souza}, {de Torres}, {del Peloso}, {del
  Pozo}, {Delbo}, {Delgado}, {Delisle}, {Demouchy}, {Dharmawardena}, {Di
  Matteo}, {Diakite}, {Diener}, {Distefano}, {Dolding}, {Edvardsson}, {Enke},
  {Fabre}, {Fabrizio}, {Faigler}, {Fedorets}, {Fernique}, {Fienga}, {Figueras},
  {Fournier}, {Fouron}, {Fragkoudi}, {Gai}, {Garcia-Gutierrez},
  {Garcia-Reinaldos}, {Garc{\'\i}a-Torres}, {Garofalo}, {Gavel}, {Gavras},
  {Gerlach}, {Geyer}, {Giacobbe}, {Gilmore}, {Girona}, {Giuffrida}, {Gomel},
  {Gomez}, {Gonz{\'a}lez-N{\'u}{\~n}ez}, {Gonz{\'a}lez-Santamar{\'\i}a},
  {Gonz{\'a}lez-Vidal}, {Granvik}, {Guillout}, {Guiraud},
  {Guti{\'e}rrez-S{\'a}nchez}, {Guy}, {Hatzidimitriou}, {Hauser}, {Haywood},
  {Helmer}, {Helmi}, {Sarmiento}, {Hidalgo}, {Hilger}, {H{\l}adczuk}, {Hobbs},
  {Holland}, {Huckle}, {Jardine}, {Jasniewicz}, {Jean-Antoine Piccolo},
  {Jim{\'e}nez-Arranz}, {Jorissen}, {Juaristi Campillo}, {Julbe}, {Karbevska},
  {Kervella}, {Khanna}, {Kontizas}, {Kordopatis}, {Korn}, {K{\'o}sp{\'a}l},
  {Kostrzewa-Rutkowska}, {Kruszy{\'n}ska}, {Kun}, {Laizeau}, {Lambert},
  {Lanza}, {Lasne}, {Le Campion}, {Lebreton}, {Lebzelter}, {Leccia}, {Leclerc},
  {Lecoeur-Taibi}, {Liao}, {Licata}, {Lindstr{\o}m}, {Lister}, {Livanou},
  {Lobel}, {Lorca}, {Loup}, {Madrero Pardo}, {Magdaleno Romeo}, {Managau},
  {Mann}, {Manteiga}, {Marchant}, {Marconi}, {Marcos}, {Marcos Santos},
  {Mar{\'\i}n Pina}, {Marinoni}, {Marocco}, {Marshall}, {Polo},
  {Mart{\'\i}n-Fleitas}, {Marton}, {Mary}, {Masip}, {Massari},
  {Mastrobuono-Battisti}, {Mazeh}, {McMillan}, {Messina}, {Michalik}, {Millar},
  {Mints}, {Molina}, {Molinaro}, {Moln{\'a}r}, {Monari}, {Mongui{\'o}},
  {Montegriffo}, {Montero}, {Mor}, {Mora}, {Morbidelli}, {Morel}, {Morris},
  {Muraveva}, {Murphy}, {Musella}, {Nagy}, {Noval}, {Oca{\~n}a}, {Ogden},
  {Ordenovic}, {Osinde}, {Pagani}, {Pagano}, {Palaversa}, {Palicio},
  {Pallas-Quintela}, {Panahi}, {Payne-Wardenaar}, {Pe{\~n}alosa Esteller},
  {Penttil{\"a}}, {Pichon}, {Piersimoni}, {Pineau}, {Plachy}, {Plum}, {Poggio},
  {Pr{\v{s}}a}, {Pulone}, {Racero}, {Ragaini}, {Rainer}, {Raiteri}, {Rambaux},
  {Ramos}, {Ramos-Lerate}, {Re Fiorentin}, {Regibo}, {Richards}, {Rios Diaz},
  {Ripepi}, {Riva}, {Rix}, {Rixon}, {Robichon}, {Robin}, {Robin}, {Roelens},
  {Rogues}, {Rohrbasser}, {Romero-G{\'o}mez}, {Rowell}, {Royer}, {Ruz Mieres},
  {Rybicki}, {Sadowski}, {S{\'a}ez N{\'u}{\~n}ez}, {Sagrist{\`a} Sell{\'e}s},
  {Sahlmann}, {Salguero}, {Samaras}, {Sanchez Gimenez}, {Sanna},
  {Santove{\~n}a}, {Sarasso}, {Schultheis}, {Sciacca}, {Segol}, {Segovia},
  {S{\'e}gransan}, {Semeux}, {Shahaf}, {Siddiqui}, {Siebert}, {Siltala},
  {Silvelo}, {Slezak}, {Slezak}, {Smart}, {Snaith}, {Solano}, {Solitro},
  {Souami}, {Souchay}, {Spagna}, {Spina}, {Spoto}, {Steele},
  {Steidelm{\"u}ller}, {Stephenson}, {S{\"u}veges}, {Surdej}, {Szabados},
  {Szegedi-Elek}, {Taris}, {Taylo}, {Teixeira}, {Tolomei}, {Tonello}, {Torra},
  {Torra}, {Torralba Elipe}, {Trabucchi}, {Tsounis}, {Turon}, {Ulla}, {Unger},
  {Vaillant}, {van Dillen}, {van Reeven}, {Vanel}, {Vecchiato}, {Viala},
  {Vicente}, {Voutsinas}, {Weiler}, {Wevers}, {Wyrzykowski}, {Yoldas}, {Yvard},
  {Zhao}, {Zorec}, {Zucker}, \& {Zwitter}}]{Gaia_2022k}
{Gaia Collaboration}, {Vallenari}, A., {Brown}, A.~G.~A., {et~al.} 2022, arXiv
  e-prints, arXiv:2208.00211, \dodoi{10.48550/arXiv.2208.00211}

\bibitem[{{Gehrz} {et~al.}(1998){Gehrz}, {Truran}, {Williams}, \&
  {Starrfield}}]{Gerhz_1998}
{Gehrz}, R.~D., {Truran}, J.~W., {Williams}, R.~E., \& {Starrfield}, S. 1998,
  \pasp, 110, 3, \dodoi{10.1086/316107}

\bibitem[{{Gray} \& {Corbally}(2009)}]{Gray_Corbally_2009}
{Gray}, R.~O., \& {Corbally}, Christopher, J. 2009, {Stellar Spectral
  Classification}

\bibitem[{{Hachisu} \& {Kato}(2004)}]{Hachisu_2004}
{Hachisu}, I., \& {Kato}, M. 2004, \apjl, 612, L57, \dodoi{10.1086/424595}

\bibitem[{{Hachisu} \& {Kato}(2010)}]{Hachisu_2010}
---. 2010, \apj, 709, 680, \dodoi{10.1088/0004-637X/709/2/680}

\bibitem[{{Hachisu} \& {Kato}(2015)}]{Hachisu_2015}
---. 2015, \apj, 798, 76, \dodoi{10.1088/0004-637X/798/2/76}

\bibitem[{{Hillier}(1990)}]{Hillier_1990}
{Hillier}, D.~J. 1990, \aap, 231, 116

\bibitem[{{Hillier} \& {Miller}(1998)}]{Hillier_1998}
{Hillier}, D.~J., \& {Miller}, D.~L. 1998, \apj, 496, 407,
  \dodoi{10.1086/305350}

\bibitem[{{Hillman} {et~al.}(2014){Hillman}, {Prialnik}, {Kovetz}, {Shara}, \&
  {Neill}}]{Hillman_2014}
{Hillman}, Y., {Prialnik}, D., {Kovetz}, A., {Shara}, M.~M., \& {Neill}, J.~D.
  2014, \mnras, 437, 1962, \dodoi{10.1093/mnras/stt2027}

\bibitem[{{Iben} \& {Tutukov}(1985)}]{Iben_1985}
{Iben}, I., J., \& {Tutukov}, A.~V. 1985, \apjs, 58, 661,
  \dodoi{10.1086/191054}

\bibitem[{{Iijima}(2006)}]{Iijima_2006}
{Iijima}, T. 2006, \aap, 451, 563, \dodoi{10.1051/0004-6361:20053984}

\bibitem[{{Iijima} {et~al.}(1998){Iijima}, {Rosino}, \& {della
  Valle}}]{Iijima_1998}
{Iijima}, T., {Rosino}, L., \& {della Valle}, M. 1998, \aap, 338, 1006

\bibitem[{{Iliadis}(2015)}]{Iliadis_2015}
{Iliadis}, C. 2015, {Nuclear Physics of Stars}, \dodoi{10.1002/9783527692668}

\bibitem[{{Jos{\'e}} \& {Hernanz}(1998)}]{Jose_1998}
{Jos{\'e}}, J., \& {Hernanz}, M. 1998, \apj, 494, 680, \dodoi{10.1086/305244}

\bibitem[{{Jos{\'e}} {et~al.}(2006){Jos{\'e}}, {Hernanz}, \&
  {Iliadis}}]{Jose_2006}
{Jos{\'e}}, J., {Hernanz}, M., \& {Iliadis}, C. 2006, \nphysa, 777, 550,
  \dodoi{10.1016/j.nuclphysa.2005.02.121}

\bibitem[{{Jos{\'e}} {et~al.}(2020){Jos{\'e}}, {Shore}, \&
  {Casanova}}]{Jose_2020}
{Jos{\'e}}, J., {Shore}, S.~N., \& {Casanova}, J. 2020, \aap, 634, A5,
  \dodoi{10.1051/0004-6361/201936893}

\bibitem[{{Kato} \& {Hachisu}(1994)}]{Kato_1994}
{Kato}, M., \& {Hachisu}, I. 1994, \apj, 437, 802, \dodoi{10.1086/175041}

\bibitem[{{Kato} {et~al.}(2022){Kato}, {Saio}, \& {Hachisu}}]{Kato_eROSITA}
{Kato}, M., {Saio}, H., \& {Hachisu}, I. 2022, \apjl, 935, L15,
  \dodoi{10.3847/2041-8213/ac85c1}

\bibitem[{{Kepler} {et~al.}(2016){Kepler}, {Koester}, \&
  {Ourique}}]{Kepler_2016}
{Kepler}, S.~O., {Koester}, D., \& {Ourique}, G. 2016, Science, 352, 67,
  \dodoi{10.1126/science.aad6705}

\bibitem[{{Kurita} {et~al.}(2020){Kurita}, {Kino}, {Iwamuro}, {Ohta}, {Nogami},
  {Izumiura}, {Yoshida}, {Matsubayashi}, {Kuroda}, {Nakatani}, {Yamamoto},
  {Tsutsui}, {Iribe}, {Jikuya}, {Ohtani}, {Shibata}, {Takahashi}, {Tokoro},
  {Maihara}, \& {Nagata}}]{Seimei}
{Kurita}, M., {Kino}, M., {Iwamuro}, F., {et~al.} 2020, \pasj, 72, 48,
  \dodoi{10.1093/pasj/psaa036}

\bibitem[{{Lauffer} {et~al.}(2018){Lauffer}, {Romero}, \&
  {Kepler}}]{Lauffer_2018}
{Lauffer}, G.~R., {Romero}, A.~D., \& {Kepler}, S.~O. 2018, \mnras, 480, 1547,
  \dodoi{10.1093/mnras/sty1925}

\bibitem[{{Lyke} \& {Campbell}(2009)}]{Lyke_2009}
{Lyke}, J.~E., \& {Campbell}, R.~D. 2009, \aj, 138, 1090,
  \dodoi{10.1088/0004-6256/138/4/1090}

\bibitem[{{Maehara} {et~al.}(2021){Maehara}, {Taguchi}, {Tampo}, {Kojiguchi},
  \& {Isogai}}]{ATel_14471}
{Maehara}, H., {Taguchi}, K., {Tampo}, Y., {Kojiguchi}, N., \& {Isogai}, K.
  2021, The Astronomer's Telegram, 14471, 1

\bibitem[{{Mahoney} {et~al.}(1982){Mahoney}, {Ling}, {Jacobson}, \&
  {Lingenfelter}}]{Mahoney_1982}
{Mahoney}, W.~A., {Ling}, J.~C., {Jacobson}, A.~S., \& {Lingenfelter}, R.~E.
  1982, \apj, 262, 742, \dodoi{10.1086/160469}

\bibitem[{{Mahoney} {et~al.}(1984){Mahoney}, {Ling}, {Wheaton}, \&
  {Jacobson}}]{Mahoney_1984}
{Mahoney}, W.~A., {Ling}, J.~C., {Wheaton}, W.~A., \& {Jacobson}, A.~S. 1984,
  \apj, 286, 578, \dodoi{10.1086/162632}

\bibitem[{{Matsubayashi} {et~al.}(2019){Matsubayashi}, {Ohta}, {Iwamuro},
  {Iwata}, {Kambe}, {Tsutsui}, {Izumiura}, {Yoshida}, \& {Hattori}}]{KOOLS-IFU}
{Matsubayashi}, K., {Ohta}, K., {Iwamuro}, F., {et~al.} 2019, \pasj, 71, 102,
  \dodoi{10.1093/pasj/psz087}

\bibitem[{{Miko{\l}ajewska} \& {Shara}(2017)}]{Mikolajewska_2017}
{Miko{\l}ajewska}, J., \& {Shara}, M.~M. 2017, \apj, 847, 99,
  \dodoi{10.3847/1538-4357/aa87b6}

\bibitem[{{Munari} \& {Valisa}(2022)}]{ATel_15796}
{Munari}, U., \& {Valisa}, P. 2022, The Astronomer's Telegram, 15796, 1

\bibitem[{{Munari} {et~al.}(2021{\natexlab{a}}){Munari}, {Valisa}, \&
  {Dallaporta}}]{ATel_14614}
{Munari}, U., {Valisa}, P., \& {Dallaporta}, S. 2021{\natexlab{a}}, The
  Astronomer's Telegram, 14614, 1

\bibitem[{{Munari} {et~al.}(2021{\natexlab{b}}){Munari}, {Valisa}, \&
  {Dallaporta}}]{ATel_14476}
---. 2021{\natexlab{b}}, The Astronomer's Telegram, 14476, 1

\bibitem[{{Noguchi} {et~al.}(2002){Noguchi}, {Aoki}, {Kawanomoto}, {Ando},
  {Honda}, {Izumiura}, {Kambe}, {Okita}, {Sadakane}, {Sato}, {Tajitsu},
  {Takada-Hidai}, {Tanaka}, {Watanabe}, \& {Yoshida}}]{Noguchi_2002}
{Noguchi}, K., {Aoki}, W., {Kawanomoto}, S., {et~al.} 2002, \pasj, 54, 855,
  \dodoi{10.1093/pasj/54.6.855}

\bibitem[{{Nomoto}(1984)}]{Nomoto_1984}
{Nomoto}, K. 1984, \apj, 277, 791, \dodoi{10.1086/161749}

\bibitem[{{Nomoto}(1987)}]{Nomoto_1987}
---. 1987, \apj, 322, 206, \dodoi{10.1086/165716}

\bibitem[{{Orio} {et~al.}(2020){Orio}, {Drake}, {Ness}, {Behar}, {Luna},
  {Darnley}, {Gallagher}, {Gehrz}, {Kuin}, {Mikolajewska}, {Ospina}, {Page},
  {Poggiani}, {Starrfield}, {Williams}, \& {Woodward}}]{Orio_2020}
{Orio}, M., {Drake}, J.~J., {Ness}, J.~U., {et~al.} 2020, \apj, 895, 80,
  \dodoi{10.3847/1538-4357/ab8c4d}

\bibitem[{{Payne-Gaposchkin}(1964)}]{Payne-Gaposchkin_1964}
{Payne-Gaposchkin}, C. 1964, {The galactic novae}

\bibitem[{{Poggiani}(2008)}]{Poggiani_2008}
{Poggiani}, R. 2008, \na, 13, 557, \dodoi{10.1016/j.newast.2008.03.001}

\bibitem[{{Politano} {et~al.}(1995){Politano}, {Starrfield}, {Truran}, {Weiss},
  \& {Sparks}}]{Politano_1995}
{Politano}, M., {Starrfield}, S., {Truran}, J.~W., {Weiss}, A., \& {Sparks},
  W.~M. 1995, \apj, 448, 807, \dodoi{10.1086/176009}

\bibitem[{{Rafanelli} \& {Rosino}(1978)}]{Rafanelli_1978}
{Rafanelli}, P., \& {Rosino}, L. 1978, \aaps, 31, 337

\bibitem[{{Ricker} {et~al.}(2015){Ricker}, {Winn}, {Vanderspek}, {Latham},
  {Bakos}, {Bean}, {Berta-Thompson}, {Brown}, {Buchhave}, {Butler}, {Butler},
  {Chaplin}, {Charbonneau}, {Christensen-Dalsgaard}, {Clampin}, {Deming},
  {Doty}, {De Lee}, {Dressing}, {Dunham}, {Endl}, {Fressin}, {Ge}, {Henning},
  {Holman}, {Howard}, {Ida}, {Jenkins}, {Jernigan}, {Johnson}, {Kaltenegger},
  {Kawai}, {Kjeldsen}, {Laughlin}, {Levine}, {Lin}, {Lissauer}, {MacQueen},
  {Marcy}, {McCullough}, {Morton}, {Narita}, {Paegert}, {Palle}, {Pepe},
  {Pepper}, {Quirrenbach}, {Rinehart}, {Sasselov}, {Sato}, {Seager},
  {Sozzetti}, {Stassun}, {Sullivan}, {Szentgyorgyi}, {Torres}, {Udry}, \&
  {Villasenor}}]{2015JATIS...1a4003R}
{Ricker}, G.~R., {Winn}, J.~N., {Vanderspek}, R., {et~al.} 2015, Journal of
  Astronomical Telescopes, Instruments, and Systems, 1, 014003,
  \dodoi{10.1117/1.JATIS.1.1.014003}

\bibitem[{{Saizar} {et~al.}(1996){Saizar}, {Pachoulakis}, {Shore},
  {Starrfield}, {Williams}, {Rothschild}, \& {Sonneborn}}]{Saizar_1996}
{Saizar}, P., {Pachoulakis}, I., {Shore}, S.~N., {et~al.} 1996, \mnras, 279,
  280, \dodoi{10.1093/mnras/279.1.280}

\bibitem[{{Saizar} {et~al.}(1992){Saizar}, {Starrfield}, {Ferland}, {Wagner},
  {Truran}, {Kenyon}, {Sparks}, {Williams}, \& {Stryker}}]{Saizar_1992}
{Saizar}, P., {Starrfield}, S., {Ferland}, G.~J., {et~al.} 1992, \apj, 398,
  651, \dodoi{10.1086/171890}

\bibitem[{{Schaefer}(2021)}]{2021arXiv210613907S}
{Schaefer}, B.~E. 2021, arXiv e-prints, arXiv:2106.13907.
\newblock \doarXiv{2106.13907}

\bibitem[{{Schwarz}(2002)}]{Schwarz_2002}
{Schwarz}, G.~J. 2002, \apj, 577, 940, \dodoi{10.1086/342234}

\bibitem[{{Schwarz} {et~al.}(2007){Schwarz}, {Shore}, {Starrfield}, \&
  {Vanlandingham}}]{Schwarz_2007}
{Schwarz}, G.~J., {Shore}, S.~N., {Starrfield}, S., \& {Vanlandingham}, K.~M.
  2007, \apj, 657, 453, \dodoi{10.1086/510661}

\bibitem[{{Shore} {et~al.}(2021{\natexlab{a}}){Shore}, {Teyssier}, {Garde},
  {Charbonnel}, {Bajer}, {Boussin}, \& {Allen}}]{ATel_14622}
{Shore}, S.~N., {Teyssier}, F., {Garde}, O., {et~al.} 2021{\natexlab{a}}, The
  Astronomer's Telegram, 14622, 1

\bibitem[{{Shore} {et~al.}(2003){Shore}, {Schwarz}, {Bond}, {Downes},
  {Starrfield}, {Evans}, {Gehrz}, {Hauschildt}, {Krautter}, \&
  {Woodward}}]{Shore_2003}
{Shore}, S.~N., {Schwarz}, G., {Bond}, H.~E., {et~al.} 2003, \aj, 125, 1507,
  \dodoi{10.1086/367803}

\bibitem[{{Shore} {et~al.}(2021{\natexlab{b}}){Shore}, {Buil}, {Dubovsky},
  {Berardi}, {Bajer}, {Boussin}, {Guarr{\`o}}, {Teyssier}, {Bertrand},
  {Viannet}, {Boyd}, {Boubault}, {Allen}, {Le Lain}, {le D{\^u}}, {Cazzato},
  {Coffin}, {Shank}, {Bressinck}, {Garde}, {Gurney}, \&
  {Leadbeater}}]{ATel_14577}
{Shore}, S.~N., {Buil}, C., {Dubovsky}, P., {et~al.} 2021{\natexlab{b}}, The
  Astronomer's Telegram, 14577, 1

\bibitem[{{Skarka} {et~al.}(2017){Skarka}, {Ma{\v{s}}ek}, {Br{\'a}t},
  {Caga{\v{s}}}, {Jury{\v{s}}ek}, {Ho{\v{n}}kov{\'a}}, {Zejda},
  {{\v{S}}melcer}, {Jel{\'\i}nek}, {Lomoz}, {Tyl{\v{s}}ar}, {Trnka}, {Pejcha},
  {Pintr}, {Lehk{\'y}}, {Jan{\'\i}k}, {{\v{C}}ervinka}, {P{\v{r}}ib{\'\i}k},
  {Motl}, {Walter}, {Zasche}, {Koss}, {H{\'a}jek}, {B{\'\i}lek}, {Li{\v{s}}ka},
  {Ku{\v{c}}{\'a}kov{\'a}}, {Bodn{\'a}r}, {Ber{\'a}nek},
  {{\v{S}}af{\'a}{\v{r}}}, {Moudr{\'a}}, {Or{\v{s}}ul{\'a}k}, {Pintr},
  {Sobotka}, {D{\v{r}}ev{\v{e}}n{\'y}}, {Jur{\'a}{\v{n}}ov{\'a}}, {Pol{\'a}k},
  {Polster}, {Onderkov{\'a}}, {Smolka}, {Auer}, {Koci{\'a}n}, {Hlad{\'\i}k},
  {Caga{\v{s}}}, {Gre{\v{s}}}, {M{\H{u}}ller}, {{\v{C}}apkov{\'a}},
  {Kysel{\'y}}, {Hornoch}, {Truparov{\'a}}, {Timko}, {Bro{\v{z}}},
  {B{\'\i}lek}, {{\v{S}}ebela}, {Han{\v{z}}l}, {{\v{Z}}ampachov{\'a}},
  {Seck{\'a}}, {Pravec}, {Mr{\v{n}}{\'a}k}, {Svoboda}, {Ehrenberger},
  {Novotn{\'y}}, {Poddan{\'y}}, {Prudil}, {Kuch{\v{t}}{\'a}k}, \&
  {{\v{S}}tegner}}]{CzeV}
{Skarka}, M., {Ma{\v{s}}ek}, M., {Br{\'a}t}, L., {et~al.} 2017, Open European
  Journal on Variable Stars, 185, 1.
\newblock \doarXiv{1709.08851}

\bibitem[{{Starrfield} {et~al.}(1972){Starrfield}, {Truran}, {Sparks}, \&
  {Kutter}}]{Starrfield_1972}
{Starrfield}, S., {Truran}, J.~W., {Sparks}, W.~M., \& {Kutter}, G.~S. 1972,
  \apj, 176, 169, \dodoi{10.1086/151619}

\bibitem[{{Strope} {et~al.}(2010){Strope}, {Schaefer}, \&
  {Henden}}]{Strope_2010}
{Strope}, R.~J., {Schaefer}, B.~E., \& {Henden}, A.~A. 2010, \aj, 140, 34,
  \dodoi{10.1088/0004-6256/140/1/34}

\bibitem[{{Taguchi} {et~al.}(2021{\natexlab{a}}){Taguchi}, {Isogai}, {Shibata},
  {Tampo}, {Kojiguchi}, \& {Maehara}}]{ATel_14478}
{Taguchi}, K., {Isogai}, K., {Shibata}, M., {et~al.} 2021{\natexlab{a}}, The
  Astronomer's Telegram, 14478, 1

\bibitem[{{Taguchi} {et~al.}(2021{\natexlab{b}}){Taguchi}, {Maehara}, {Isogai},
  {Tampo}, {Kojiguchi}, {Kato}, \& {Nogami}}]{ATel_14472}
{Taguchi}, K., {Maehara}, H., {Isogai}, K., {et~al.} 2021{\natexlab{b}}, The
  Astronomer's Telegram, 14472, 1

\bibitem[{{Tajitsu} {et~al.}(2010){Tajitsu}, {Aoki}, {Kawanomoto}, \&
  {Narita}}]{Tajitsu_2010}
{Tajitsu}, A., {Aoki}, W., {Kawanomoto}, S., \& {Narita}, N. 2010, Publications
  of the National Astronomical Observatory of Japan, 13, 1

\bibitem[{{Tajitsu} {et~al.}(2015){Tajitsu}, {Sadakane}, {Naito}, {Arai}, \&
  {Aoki}}]{Tajitsu_2015}
{Tajitsu}, A., {Sadakane}, K., {Naito}, H., {Arai}, A., \& {Aoki}, W. 2015,
  \nat, 518, 381, \dodoi{10.1038/nature14161}

\bibitem[{{Takeda} {et~al.}(2018){Takeda}, {Diaz}, {Campbell}, \&
  {Lyke}}]{Takeda_2018}
{Takeda}, L., {Diaz}, M., {Campbell}, R., \& {Lyke}, J. 2018, \mnras, 473, 355,
  \dodoi{10.1093/mnras/stx2285}

\bibitem[{{Tanaka} {et~al.}(2011){Tanaka}, {Nogami}, {Fujii}, {Ayani}, {Kato},
  {Maehara}, {Kiyota}, \& {Nakajima}}]{Tanaka_2011}
{Tanaka}, J., {Nogami}, D., {Fujii}, M., {et~al.} 2011, \pasj, 63, 911,
  \dodoi{10.1093/pasj/63.4.911}

\bibitem[{{Vanlandingham} {et~al.}(1997){Vanlandingham}, {Starrfield}, \&
  {Shore}}]{Vanlandingham_1997}
{Vanlandingham}, K.~M., {Starrfield}, S., \& {Shore}, S.~N. 1997, \mnras, 290,
  87, \dodoi{10.1093/mnras/290.1.87}

\bibitem[{{Vanlandingham} {et~al.}(1999){Vanlandingham}, {Starrfield}, {Shore},
  \& {Sonneborn}}]{Vanlandingham_1999}
{Vanlandingham}, K.~M., {Starrfield}, S., {Shore}, S.~N., \& {Sonneborn}, G.
  1999, \mnras, 308, 577, \dodoi{10.1046/j.1365-8711.1999.02731.x}

\bibitem[{{Vasini} {et~al.}(2023){Vasini}, {Matteucci}, {Spitoni}, \&
  {Siegert}}]{Vasini_2023}
{Vasini}, A., {Matteucci}, F., {Spitoni}, E., \& {Siegert}, T. 2023, \mnras,
  523, 1153, \dodoi{10.1093/mnras/stad1440}

\bibitem[{{Whitney} \& {Clayton}(1989)}]{Whitney_1989}
{Whitney}, B.~A., \& {Clayton}, G.~C. 1989, \aj, 98, 297,
  \dodoi{10.1086/115146}

\bibitem[{{Williams}(2012)}]{Williams_2012}
{Williams}, R. 2012, \aj, 144, 98, \dodoi{10.1088/0004-6256/144/4/98}

\bibitem[{{Williams}(1992)}]{Williams_1992}
{Williams}, R.~E. 1992, \aj, 104, 725, \dodoi{10.1086/116268}

\bibitem[{{Williams} {et~al.}(1994){Williams}, {Phillips}, \&
  {Hamuy}}]{Williams_1994}
{Williams}, R.~E., {Phillips}, M.~M., \& {Hamuy}, M. 1994, \apjs, 90, 297,
  \dodoi{10.1086/191864}

\bibitem[{{Yamanaka} {et~al.}(2010){Yamanaka}, {Uemura}, {Kawabata}, {Fujii},
  {Tanabe}, {Imamura}, {Komatsu}, {Arai}, {Sasada}, {Itoh}, {Harao},
  {Kunitomi}, {Nagae}, {Nose}, {Ohsugi}, {Okushima}, {Sakimoto}, \&
  {Yoshida}}]{Yamanaka_2010}
{Yamanaka}, M., {Uemura}, M., {Kawabata}, K.~S., {et~al.} 2010, \pasj, 62, L37,
  \dodoi{10.1093/pasj/62.5.L37}

\bibitem[{{Yaron} {et~al.}(2005){Yaron}, {Prialnik}, {Shara}, \&
  {Kovetz}}]{Yaron_2005}
{Yaron}, O., {Prialnik}, D., {Shara}, M.~M., \& {Kovetz}, A. 2005, \apj, 623,
  398, \dodoi{10.1086/428435}

\end{thebibliography}
\end{document}